\documentclass[aps,prd,
twocolumn,%
superscriptaddress,
showpacs,nofootinbib,eqsecnum,amsmath,amssymb,floatfix,
]{revtex4}

\usepackage{bm}
\usepackage{graphicx}
\usepackage[usenames]{color}
\usepackage{ulem}

\allowdisplaybreaks

\newcommand{\ui}{\mathrm{i}}
\newcommand{\ud}{\mathrm{d}}
\newcommand{\uD}{\mathrm{D}}

\newcommand{\IAP}{\affiliation{Institut d'Astrophysique de Paris,
   UMR 7095 CNRS Universit\'e Pierre \& Marie Curie, 98$^{\text{bis}}$
   boulevard Arago, 75014 Paris, France}}
\newcommand{\Maryland}{\affiliation{Maryland Center for Fundamental
    Physics \& Joint Space-Science Center,\\ Department of Physics, 
University of Maryland, College
    Park, MD 20742, USA}}

\begin{document}

\title{Tail-induced spin-orbit effect in the gravitational radiation of
  compact binaries}

\author{Luc Blanchet} \IAP %
\author{Alessandra Buonanno} \Maryland %
\author{Guillaume Faye} \IAP %

\begin{abstract}
  Gravitational waves contain tail effects which are due to the
  back-scattering of linear waves in the curved space-time geometry around the
  source. In this paper we improve the knowledge and accuracy of the two-body
  inspiraling post-Newtonian (PN) dynamics and gravitational-wave signal by
  computing the spin-orbit terms induced by tail effects. Notably, we derive
  those terms at 3PN order in the gravitational-wave energy flux, and 2.5PN
  and 3PN orders in the wave polarizations. This is then used to
  derive the spin-orbit tail effects in the phasing through 3PN
  order. Our results can be employed to carry out more accurate comparisons
  with numerical-relativity simulations and to improve the accuracy of
  analytical templates aimed at describing the whole process of inspiral, merger
  and ringdown.
\end{abstract}


\maketitle

\section{Introduction}
\label{sec:intro}

\subsection{Motivation}

During the last ten years a network of ground-based laser-interferometer
gravitational-wave detectors has been built and has taken data at design
sensitivity. It is a worldwide network composed of the Laser Interferometer
Gravitational wave Observatory (LIGO), Virgo, GEO-600, and TAMA and it has
operated in the frequency range $10 \mbox{--} 10^3$ Hz. Coalescing binary
systems composed of black holes and/or neutron stars are among the most
promising sources for those detectors. By $2016$ the gravitational-wave
detectors will be upgraded to a sensitivity such that event rates for
coalescing binary systems will increase by approximately a factor one
thousand, thus making likely the first detection of gravitational waves from
those systems. In the future, space-based detectors like LISA should detect
supermassive black-hole binary systems in the low frequency band $10^{-4}
\mbox{--} 10^{-2}$ Hz.

The search for gravitational waves from coalescing binary systems and the
extraction of source parameters are based on the matched-filtering technique,
which requires a rather accurate knowledge of the waveform of the incoming
signal. In particular, the detection and subsequent data analysis are made by
using a bank of templates modeling the gravitational wave emitted by the
source. The need of a faithful template bank has driven the development of
accurate templates over the last thirty years.

The post-Newtonian (PN) expansion is the most powerful approximation scheme in
analytical relativity capable of describing the two-body dynamics and
gravitational-wave emission of inspiraling compact binary
systems~\cite{Bliving}. The PN expansion is an expansion in the ratio of the
characteristic orbital velocity of the binary $v$ to the speed of light
$c$. However, as the black holes approach each other toward merger, we expect
the PN expansion to lose accuracy because the velocity of the holes approaches
the speed of light. At that point, numerical relativity
(NR)~\cite{Pretorius,Campanelli,Baker} plays a crucial role providing us with
the dynamics and gravitational-wave emission of the last cycles of inspiral,
followed by the merger and ringdown phases. Furthermore, by properly combining
PN predictions and NR results, it is possible to describe analytically and/or
numerically with high accuracy, the full gravitational-wave
signal~\cite{DN09,Buo09,Pan09,Ajith08}.

Black holes in binary systems can carry spin, and when spins are not aligned
with the orbital angular momentum, spins induce precession of the orbital
plane (see e.g. Ref.~\cite{ACST94}). This adds substantial complexity to the
gravitational waveforms, making it indispensable to include spin effects in
templates used for the search. Moreover, as found long time
ago~\cite{BoR66,CTJN67,HR69,BD86,BD88,BD92,BS93,AF97,B98tail}, gravitational
waves contain tail effects which are due to the back-scattering of linear
waves in the curved space-time geometry around the source (and primarily
generated by its mass). This causes the gravitational-wave signal to depend on
the entire history of the binary system.

In this paper we improve the knowledge and accuracy of the two-body
inspiraling dynamics and gravitational-wave signal by computing the spin-orbit
(SO) terms induced by tail effects. This is the continuation of our previous
work on spins~\cite{FBB06,BBF06} where we obtained the next-to-leading 2.5PN
SO contributions in the equations of motion and gravitational-wave energy
flux. Here, we derive those SO terms at 3PN order in the gravitational-wave
energy flux, where they are entirely due to tails. Furthermore we obtain the
SO terms at 2.5PN and 3PN orders in the wave polarizations that are
specifically due to tails, leaving aside other SO terms at these orders that
come from instantaneous (non-tail) linear contributions and which will not be
computed here.

We obtain the energy flux in two independent ways, first directly using the
radiative multipole moments, and second by differentiating and squaring the
gravitational-wave polarizations. To compute the SO tail effects in the wave
polarizations we solve the two-body dynamics taking into account spin
precession. Assuming quasi-circular adiabatic inspiral, we also compute the
3PN SO terms induced by tails in the gravitational-wave phasing. These results
can be used to improve the accuracy of inspiraling templates, to carry out
comparison with numerical-relativity predictions, and to improve the accuracy
of effective-one-body and phenomenological
templates~\cite{DN09,Buo09,Pan09,Ajith08}.

As an important check of our results we obtain the 3PN SO tail terms in the
energy flux in the test-particle limit, and find perfect agreement with
earlier PN computations based on black-hole perturbation induced by the motion
of a test particle around a massive black hole~\cite{TSTS96}.

This paper is organized as follows. In Sec.~\ref{sec:multipole} we review the
post-Newtonian multipole moment formalism and discuss relevant properties of
tails. In Sec.~\ref{sec:appl}, we describe how spin effects are included in
the PN formalism and derive the binary's evolution equations when black holes
carry spins. In Sec.~\ref{sec:nlambdaell} we obtain the time evolution of the
moving triad and solve the precessing dynamics at the relevant PN order. In
Sec.~\ref{sec:comput} we compute the 2.5PN and 3PN SO tail effects in the
gravitational waveform and polarizations. Restricting ourselves to
quasi-circular adiabatic inspiral, we derive in Sec.~\ref{sec:Eflux} the 3PN
SO tail effects in the energy flux and in the gravitational
phasing. Section~\ref{sec:concl} contains our main conclusions. In
Appendix~\ref{sec:eomcheck} we check that the equations of motion do not
contain physical SO effects at 3PN order. In Appendix~\ref{sec:secular} we
investigate the secular evolution of the spins, and finally in
Appendix~\ref{sec:polarizations} we give the explicit gravitational-wave
polarizations.

\subsection{Notation}
\label{sec:notation}

In this paper we adopt the following notations. $G$ is the Newton constant and
$c$ is the speed of light. As usual we refer to $n$PN as the post-Newtonian
terms with formal order $\mathcal{O}(c^{-2n})$ relative to the Newtonian
acceleration in the equations of motion, or to the lowest-order
quadrupole-moment formalism for the radiation field. Greek indices are
space-time indices, and Latin are space indices. The quantity $L=i_1\cdots
i_\ell$ denotes a multi-index composed of $\ell$ multipolar spatial indices
$i_1, \cdots, i_\ell$ ranging from 1 to 3. Similarly $L-1=i_1\cdots
i_{\ell-1}$ and $kL-2=k i_1\cdots i_{\ell-2}$; $N_L = N_{i_1}\cdots
N_{i_\ell}$ is the product of $\ell$ spatial vectors $N_i$ (similarly for $x_L
= x_{i_1}\cdots x_{i_\ell}$). We indicate with $\partial_L
=\partial_{i_1}\cdots \partial_{i_\ell}$ and $\partial_{kL-2} =
\partial_{k}\partial_{i_1}\cdots \partial_{i_{\ell-2}}$ the product of partial
derivatives $\partial_i=\partial/\partial x^i$. In the case of summed-up
(dummy) multi-indices $L$, we do not write the $\ell$ summations from 1 to 3
over their indices. Furthermore, the symmetric-trace-free (STF) projection is
indicated using brackets, $T_{\langle L\rangle}=\mathrm{STF}[T_L]$; thus
$U_L=U_{\langle L\rangle}$ and $V_L=V_{\langle L\rangle}$ for STF moments. For
instance we write $x_{\langle i}v_{j\rangle}=\frac{1}{2}(x_iv_j+x_jv_i)
-\frac{1}{3}\delta_{ij}\bm{x}\cdot\bm{v}$. The transverse-traceless (TT)
projection operator is denoted $\mathcal{P}^\mathrm{TT}_{ijkl} =
\mathcal{P}_{ik}\mathcal{P}_{jl}-\frac{1}{2}\mathcal{P}_{ij}\mathcal{P}_{kl}$
where $\mathcal{P}_{ij}=\delta_{ij}-N_iN_j$ is the projector orthogonal to the
unit direction $\bm{N}=\bm{X}/R$ of a radiative coordinate system
$X^\mu=(c\,T,\bm{X})$. The quantity $\varepsilon_{ijk}$ is the Levi-Civita
antisymmetric symbol such that $\varepsilon_{123}=1$. Finally, we denote time
derivatives with a superscript $(n)$, and we indicate the symmetrization
operation with round parentheses.

\section{Wave generation formalism}
\label{sec:multipole}

The gravitational waveform $h_{ij}^\mathrm{TT}$, generated by an isolated
source described by a stress-energy tensor $T^{\mu\nu}$ with compact
support, and propagating in the asymptotic regions of the source, is the TT
projection of the metric deviation at the leading-order $1/R$ in the distance
to the source. It is parametrized by STF mass-type
moments $U_L$ and current-type ones $V_L$, which constitute the observables of
the waveform at infinity from the source and are called the radiative moments
\cite{Th80}. The general expression of the TT waveform, in a suitable
radiative coordinate system $X^\mu=(c\,T,\bm{X})$, reads, when neglecting
terms of the order of $1/R^2$ or higher,
\begin{align}\label{eq:hijTT}
h^\mathrm{TT}_{ij} &= \frac{4G}{c^2R} \,\mathcal{P}^\mathrm{TT}_{ijkl}
\sum^{+\infty}_{\ell=2}\frac{N_{L-2}}{c^\ell\ell !} \biggl[
U_{klL-2} \nonumber \\ &\qquad\qquad- \frac{2\ell}{c(\ell+1)}\,N_{m}
\,\varepsilon_{mn(k}
\,V_{l)nL-2} \biggr]\,.
\end{align}
Here the radiative moments $U_L$ and $V_L$ are functions of the retarded time
$T_R\equiv T-R/c$ in the radiative coordinate system (we denote
$R=\vert\bm{X}\vert$). The integer $\ell$ refers to the multipolar order, and
$\bm{N} = \bm{X}/R = (N_i)$ is the unit vector pointing from the source to the
far away detector. The TT projection operator $\mathcal{P}^\mathrm{TT}_{ijkl}$
and other notations are defined in Sec.~\ref{sec:notation}. With $\bm{P}=
(P_i)$ and $\bm{Q}= (Q_i)$ denoting two unit polarization vectors, orthogonal
and transverse to the direction of propagation $\bm{N}$,
the two ``plus'' and ``cross'' polarization states of the waveform
read as 
\begin{subequations}
\label{eq:hp}
\begin{align}\label{eq:hpc}
h_+ &= \frac{P_iP_j-Q_iQ_j}{2} \,h^\mathrm{TT}_{ij}\,,\\
\label{eq:hpp}
h_\times &= \frac{P_iQ_j+P_jQ_i}{2} \,h^\mathrm{TT}_{ij}\,.
\end{align}\end{subequations}
Our convention for the choice of the polarization vectors $\bm{P}$ and
$\bm{Q}$ in the case of binary systems will be specified in
Fig.~\ref{figure:SourceFrame}.  Plugging Eq.~\eqref{eq:hijTT} into the
standard expression for the gravitational-wave energy flux we get~\cite{Th80}
\begin{multline}
\mathcal{F} = \sum_{\ell = 2}^{+ \infty} \frac{G}{c^{2\ell +1}}\,\biggl[
\frac{(\ell+1)(\ell+2)}{(\ell-1) \ell \, \ell! (2\ell+1)!!} U_L^{(1)} U_L^{(1)} \\ 
+ \frac{4\ell (\ell+2)}{c^2 (\ell-1) (\ell+1)!
  (2\ell+1)!!} V_L^{(1)} V_L^{(1)}\biggr]\,.
\end{multline}

\subsection{Expression of the radiative moments}
\label{sec:radmom}

In the multipolar-post-Minkowskian formalism~\cite{BD86,BD88,BD92}, the
radiative moments are expressed in terms of two other sets of moments,
referred to as the ``canonical'' moments $M_L$, $S_L$, and which are relevant
to the description of the source's near zone. The relation between the
radiative moments $U_L$, $V_L$ and the canonical ones $M_L$, $S_L$ encodes all
the non-linearities in the wave propagation between the source and the
detector~\cite{BD92}. Those relations may be re-expanded in a PN way and are
then seen to contain, at the leading 1.5PN order, the contribution of the
so-called gravitational-wave tails, due to backscattering of linear waves onto
the space-time curvature associated with the total mass of the source
itself. The explicit expressions at 1.5PN order are~\cite{BD92,B95}
\begin{widetext}
\begin{subequations} \label{eq:tails}
\begin{align}
U_L(T_R) &= M_L^{(\ell)} + \frac{2 G M}{c^3} \int_{-\infty}^{T_R}\! \ud t \,
M_L^{(\ell +2)}(t) \biggl[\ln \biggl(\frac{T_R-t}{2\tau_0} \biggr)+
  \kappa_\ell \biggr] 
+\mathcal{O}\Bigl(\frac{1}{c^3}\Bigr)_\text{non-tail}\, ,\label{eq:tailsU}\\
V_L(T_R) &= 
S_L^{(\ell)} + \frac{2 G M}{c^3} \int_{-\infty}^{T_R} \!  \ud t \,
S_L^{(\ell +2)}(t) \biggl[\ln \biggl(\frac{T_R-t}{2\tau_0} \biggr)+
  \pi_\ell \biggr]
+\mathcal{O}\Bigl(\frac{1}{c^3}\Bigr)_\text{non-tail}\, ,\label{eq:tailsV}
\end{align}
\end{subequations}
\end{widetext}
where $M$ is the Arnowitt-Deser-Misner (ADM) mass associated with the source.
It also coincides with the mass monopole moment. The remainders
$\mathcal{O}(c^{-3})_\text{non-tail}$ in Eqs.~(\ref{eq:tails}) denote some
correction terms which are at least of order 1.5PN and are instantaneous or
contain the non-linear memory effect which will not be considered in the present
computation. Here $\kappa_\ell$ and $\pi_\ell$ denote some numerical
constants given by~\cite{B95}
\begin{subequations}\label{kappapi}
\begin{align}
\kappa_\ell &= {2\ell^2 +5\ell+4\over \ell(\ell+1)(\ell+2)} +
 \sum^{\ell-2}_{k=1} {1\over k} \,,\\
 \pi_\ell &= {\ell-1\over \ell(\ell+1)} +
 \sum^{\ell-1}_{k=1} {1\over k} \,.
\end{align}
\end{subequations} 
The constant $\tau_0$ in Eqs.~(\ref{eq:tails}) is a freely specifiable time
scale entering the relation between the radiative time $T_R$ and the
corresponding retarded time in harmonic coordinates.

The canonical moments $M_L$ and $S_L$ are themselves linked to six sets of
multipole moments characterizing the source, collectively called the
``source'' moments and denoted $I_L$, $J_L$, $W_L$, $X_L$, $Y_L$, $Z_L$. The
point is that those source moments are known as explicit integrals extending
over the pseudo-energy-momentum tensor of the matter fields and the
gravitational field of the source~\cite{B95,B98mult}. In the following we
shall essentially need $I_L$ and $J_L$ which represent the main mass and
current moments of the source. The other moments $W_L$, $X_L$, $Y_L$ and $Z_L$
play a little role because they are associated with a possible gauge
transformation performed at linear order.  It turns out that the difference
between the canonical moments $M_L$, $S_L$ and the source moments $I_L$, $J_L$
arises only at the 2.5PN order:
\begin{subequations}\label{eq:MLSL}
\begin{align}
M_L = I_L 
+\mathcal{O}\Bigl(\frac{1}{c^5}\Bigr)\, , \label{eq:ML}\\
S_L = J_L 
+\mathcal{O}\Bigl(\frac{1}{c^5}\Bigr)\, . \label{eq:SL}
\end{align}
\end{subequations}

For the present application it will be sufficient to consider the source mass
moments $I_L$ at 1PN order and the current ones $J_L$ at Newtonian
order (see the discussion in Sec.~\ref{sec:multspin}). These are given
by~\cite{BD89}
\begin{subequations}\label{eq:ILJLN}
  \begin{align} \label{eq:ILS} I_{L} &= \int \ud^3\bm{x}\, \biggl[
    \hat{x}_L \sigma +
    \frac{1}{2c^2(2\ell+3)}\,\hat{x}_{L}|\bm{x}|^2\,\sigma^{(2)}
    \nonumber\\&\qquad - \frac{4(2\ell+1)}{c^2(\ell+1)(2\ell+3)}\,\hat{x}_{iL}
    \,\sigma^{(1)}_{i}\biggr]+\mathcal{O}\Bigl(\frac{1}{c^4}\Bigr)\,,\\ 
\label{eq:JLS}
    J_{L} &= \varepsilon_{ij\langle i_\ell} \int \ud^3 \bm{x} \,
    \hat{x}_{L-1\rangle i} \,\sigma_{j}
    +\mathcal{O}\Bigl(\frac{1}{c^2}\Bigr)\,.
\end{align}
\end{subequations}
The other moments we shall need are the mass monopole moment $M$ 
and the monopole of the moment $W_L$, which are
given by
\begin{subequations}\label{eq:MW}
\begin{align}
M &= \int \! \ud^3 \bm{x}\, \sigma
+\mathcal{O}\Bigl(\frac{1}{c^2}\Bigr)\, ,\label{eq:M}\\
W &= \frac{1}{3} \int \! \ud^3 \bm{x} \, x^i \sigma_i
+\mathcal{O}\Bigl(\frac{1}{c^2}\Bigr)\,.\label{eq:W}
\end{align}\end{subequations}
The mass, current and tensor densities $\sigma$,
$\sigma_i$, $\sigma_{ij}$ in Eqs.~\eqref{eq:ILJLN}--\eqref{eq:MW}, are defined
as (where $T^{ii}\equiv\delta_{ij}T^{ij}$)
\begin{align}\label{eq:densities}
\sigma &= \frac{T^{00} + T^{ii}}{c^2}\,, &
\sigma_i &= \frac{T^{0i}}{c}\,, &
\sigma_{ij} &= T^{ij} \,.
\end{align}
We recall that, e.g., $\sigma^{(n)}$ 
in Eqs.~\eqref{eq:ILJLN} means taking $n$-time derivatives. 

The spin parts of the source moments in Eqs.~\eqref{eq:ILJLN} will come from
the model we adopt for the stress-energy tensor $T^{\mu\nu}$ appropriate to
spinning compact binaries (see details in Sec.~\ref{sec:appl}). Importantly,
we notice that for the accuracy required by our calculation of the spin
effects due to tails all integrands in Eqs.~\eqref{eq:ILJLN} have compact
support. This is in contrast with the spin effects at 2.5PN order which
necessitate non-compact supported higher-order terms in the source
moments~\cite{BBF06}. (We also find that the second term in $I_L$ can be
ignored in the present application to spins.)

\subsection{Computing the tail integrals}
\label{sec:tailint}

The tail integrals in Eqs.~\eqref{eq:tails} extend over the entire past of the
evolving source and it is \textit{a priori} a non trivial task to compute
them. Here we recall, based on Refs.~\cite{BD92,BS93}, that the tails are
actually very weakly sensitive (in a post-Newtonian sense) to the past history
of the source, and can essentially be computed using the \textit{current}
dynamics, i.e. at current time $T_R$, of the source.

We have to compute, e.g., the integral appearing in the radiative mass multipole
moment \eqref{eq:tailsU}, in which we can replace, following \eqref{eq:ML},
the canonical moment $M_L$ by the source moment $I_L$. Thus,
\begin{equation}\label{tailint}
\mathcal{U}_{L}(T_R) = \int_{-\infty}^{T_R} \ud t
\,I^{(\ell+2)}_{L}(t)\,\ln \biggl(\frac{T_R-t}{2\tau'_0}\biggr)\,,
\end{equation}
where we pose $\tau'_0=\tau_0 e^{-\kappa_\ell}$.

Let us introduce a constant time interval $\mathcal{T}$ to split the integral
\eqref{tailint} into some contribution coming from the ``recent past'', and
extending from the current time $T_R$ to $T_R-\mathcal{T}$, and the remaining
contribution called the ``remote past'', from $T_R-\mathcal{T}$ to $-\infty$
in the past. The recent past can be thought of as corresponding to the most
recent orbital period of a compact binary system, while the remote past will
include the details (eventually unknown) of the formation and early past
evolution of the compact binary. However we shall prove that our result is
independent of the chosen time scale $\mathcal{T}$.

To control the convergence of the tail integral \eqref{tailint} in the past we
make a physical assumption regarding the behavior of the multipole moment
$I_{L}(t)$ when $t\to-\infty$. We assume that at very early times the source
was formed from a bunch of freely falling particles initially moving on some
hyperbolic-like orbits, and forming at a later time a gravitationally bound
system by emission of gravitational radiation. The gravitational motion of
initially free particles is given by $x^i(t)=V^it+W^i\ln(-t)+X^i+o(1)$, where
$V^i$ and $X^i$ denote constant vectors, and $W^i=G m V^i/V^3$ (see
Ref.~\cite{Eder} for a proof; to simplify we consider the relative motion of
two particles with total mass $m$). Here the Landau remainder $o$-symbol
satisfies $\partial^n o(1)/\partial t^n=o(1/t^n)$. From that physical
assumption we find that the multipole moment behaves when $t\to-\infty$ like
\begin{equation}\label{Ipast}
I_{L}(t) = A_{L}t^\ell+B_{L}t^{\ell-1}\ln(-t) + C_{L} t^{\ell-1} + o(t^{\ell-1})\,,
\end{equation}
where $A_{L}$, $B_{L}$ and $C_{L}$ are constant tensors. The time derivatives
of the moment appearing in Eq.~\eqref{tailint} are therefore dominantly like
\begin{equation}\label{Ipast'}
I^{(\ell+2)}_{L}(t) = D_{L}t^{-3}+ o(t^{-3})\,,
\end{equation}
which ensures that the integral \eqref{tailint} is convergent.

Next we integrate the ``remote-past'' integral (from $T_R-\mathcal{T}$ to
$-\infty$) by parts and make use of our assumption
\eqref{Ipast}--\eqref{Ipast'} to arrive at (posing $t=T_R-\mathcal{T}x$)
\begin{eqnarray}
\label{tailsplit}
\mathcal{U}_{L}(T_R) &=& I^{(\ell+1)}_{L}(T_R)\,\ln
\biggl(\frac{\mathcal{T}}{2\tau'_0}\biggr) \nonumber \\ && +
\mathcal{T}\int_0^{1} \ud x \,\ln x\,I^{(\ell+2)}_{L}(T_R-\mathcal{T}x)
\nonumber \\ && + \int_{1}^{+\infty}\frac{\ud
  x}{x}\,I^{(\ell+1)}_{L}(T_R-\mathcal{T}x)\,.
\end{eqnarray}
At this stage it is convenient to perform a Fourier decomposition of the
multipole moment, i.e.
\begin{equation}\label{Fourier}
I_{L}(t) = \int_{-\infty}^{+\infty} \frac{\ud
  \Omega}{2\pi}\,\widetilde{I}_{L}(\Omega)\,e^{-\ui \Omega t}\,.
\end{equation}
(The Fourier coefficients satisfy
$\widetilde{I}^*_{L}(\Omega)=\widetilde{I}_{L}(-\Omega)$ since the moment is
real.) Inserting \eqref{Fourier} into \eqref{tailsplit} we obtain a
closed-form result in the Fourier domain thanks to the mathematical
formula~\cite{GR}
\begin{multline}\label{GR}
\lambda\!\int_0^{1} \ud x \ln x \,e^{\ui \lambda x} + \ui
\int_{1}^{+\infty}\frac{\ud x}{x}\,e^{\ui \lambda x} \\= - \frac{\pi}{2}
\text{s}(\lambda) - \ui \Bigl(\ln\vert\lambda\vert+\gamma_\text{E}\Bigr)\,,
\end{multline}
where $\lambda=\Omega\mathcal{T}$, with $\text{s}(\lambda)$ and
$\vert\lambda\vert$ denoting the sign and the absolute value, and
$\gamma_\text{E}$ being the Euler constant. Finally the result reads
\begin{align}\label{tailsplitres}
\mathcal{U}_{L}(T_R) &= \ui\int_{-\infty}^{+\infty} \frac{\ud
  \Omega}{2\pi}\,(-\ui\Omega)^{\ell+1}\,\widetilde{I}_{L}(\Omega)\,e^{-\ui\,\Omega\,T_R}
\nonumber\\&\qquad\times\left[\frac{\pi}{2}\text{s}(\Omega) +
  \ui\Bigl(\ln(2\vert\Omega\vert\tau'_0)+\gamma_\text{E}\Bigr)\right]\,.
\end{align}
We observe that the arbitrary time scale $\mathcal{T}$ has cancelled from this
result. 

Later we shall apply this result to the computation of the waveform and energy
flux of a spinning compact binary system. \textit{A priori}, since the tail
integral \eqref{tailint} depends on all the past history of the binary (with
the current binary's dynamics being the result of its long evolution by
gravitational radiation emission), we expect that the binary's continuous
spectrum of frequencies $\Omega$ should contain all orbital frequencies at any
epoch in the past, say $\omega(t)$ with $t\leq T_R$, besides the current
orbital frequency $\omega(T_R)$. However, it has been shown in the Appendix of
Ref.~\cite{BS93} that one can actually compute the tail integral by
considering only the \textit{current} frequency $\omega(T_R)$. Indeed the
error made by this procedure is small in a post-Newtonian sense, being of the
order of $\mathcal{O}(\xi\ln\xi)$, where $\xi=\dot{\omega}/\omega^2$ denotes
the adiabatic parameter associated with the gravitational radiation emission,
and evaluated at the current time $T_R$. In a PN expansion we have
$\xi(T_R)=\mathcal{O}(1/c^{5})$ so the error made by replacing the past
dynamics by the current one is of the order of $\mathcal{O}(\ln c/c^5)$ and
can be neglected. The proof given in Ref.~\cite{BS93} is based on a simple
model of binary evolution in the past, where an always circular orbit is
decaying by radiation following the lowest order quadrupole formula, and spins
are neglected. In this paper we shall assume that this result remains valid
for spinning binaries.

\section{Applications to spinning binaries}
\label{sec:appl}

\subsection{Spin vectors for point-like objects}\label{sec:spins}

Following our previous work~\cite{FBB06,BBF06} we base our calculations on the
model of point-particles with spins~\cite{Papa51, Papa51spin, CPapa51spin,
  BOC75, BOC79, Tulc1, Tulc2, Traut58, Dixon, BI80, D82, MST96, TMSS96}. The
stress-energy tensor $T^{\mu\nu}$ of a system of spinning particles is the sum
of a monopolar piece, made of Dirac delta-functions, plus the dipolar or spin
piece, made of gradients of delta-functions:
\begin{align}\label{eq:Tmunu}
T^{\mu\nu} =&
c^2 \sum_A
\int_{-\infty}^{+\infty} \! \ud\tau_A \biggl\{ p_A^{(\mu} u_A^{\nu)} 
\frac{\delta^{(4)}
(x-y_A)}{\sqrt{-g_A}} \nonumber \\&\quad
-\frac{1}{c}\,\nabla_\rho\biggl[ \bar{S}_A^{\rho(\mu}\,u_A^{\nu)} \frac{\delta^{(4)}
(x-y_A)}{\sqrt{-g_A}} \biggr] \biggr\} \,,
\end{align}
where $\delta^{(4)}$ is the four-dimensional Dirac function, $x^\mu$ is the
field point, $y_A^\mu$ is the world-line of particle $A$, $u_A^\mu=\ud
y_A^\mu/(c \ud\tau_A)$ is the four-velocity, such that $g^A_{\mu\nu}u_A^\mu
u_A^\nu=-1$ where $g^A_{\mu\nu}\equiv g_{\mu\nu}(y_A)$ denotes the metric at
the particle's location, $p_A^\mu$ is the linear momentum of the particle, and
$\bar{S}_A^{\mu\nu}$ denotes its antisymmetric spin angular momentum. Our
notation and conventions are the same as in Refs.~\cite{FBB06,BBF06} which
provide more details, except that here we shall denote using an overbar
(i.e. $\bar{S}_A^{\mu\nu}$) the original spin variable used in
\cite{FBB06,BBF06}. Note that with our convention the spin variable has the
dimension of an angular momentum times $c$.

In order to fix unphysical degrees of freedom associated with an arbitrariness
in the definition of $\bar{S}^{\mu\nu}$ in the case of point particles (and
associated with the freedom in the choice for the location of the
center-of-mass worldline of extended bodies), we adopt the covariant
supplementary spin condition also called Tulczyjew
condition~\cite{Tulc1,Tulc2}:
\begin{equation}\label{eq:SSC}
\bar{S}_A^{\mu\nu}\,p^A_\nu = 0 \, ,
\end{equation}
which allows the natural definition of the spin four-vector $\bar{S}^A_\mu$ in
such a way that
\begin{equation}\label{eq:spin4}
\bar{S}_A^{\mu\nu}=-\frac{1}{\sqrt{-g_A}}\,\varepsilon^{\mu\nu\rho\sigma}\,
\frac{p^A_\rho}{m_A  c} \,\bar{S}^A_\sigma \, ,
\end{equation}
where $\varepsilon^{\mu\nu\rho\sigma}$ is the four-dimensional antisymmetric
Levi-Civita symbol (such that $\varepsilon^{0123}=1$). For the spin vector
$\bar{S}^A_\mu$ itself, we choose a four-vector that is purely spatial in the
particle's instantaneous rest frame, which means that in any frame
\begin{equation}\label{eq:Su}
\bar{S}^A_\mu u_A^\mu=0 \, .
\end{equation}
This choice is also adopted in Refs.~\cite{KWW93, K95, OTO98, TOO01}. As a
consequence of the condition~\eqref{eq:SSC}, we can check that the mass
defined by $m_A^2 c^2 = -p_A^\mu p^A_\mu$ is constant along the trajectories,
i.e. $\ud m_A/\ud\tau_A=0$.

Important simplifications occur in the case of SO interactions, which are
linear in the spins. Neglecting quadratic (spin-spin) interactions, the linear
momentum is simply linked to the four velocity as $p_A^\mu=m_A c
u_A^\mu$, so the supplementary spin condition \eqref{eq:SSC} reduces to
$\bar{S}_A^{\mu\nu}\,u^A_\nu = 0$, and the equation of evolution of the spins is
given by
\begin{equation}\label{eq:paralell}
\frac{\uD \bar{S}^A_\mu}{\ud \tau_A}=0 \, ,
\end{equation}
which means that the spin is parallely transported along the particle's
trajectory.

Following \cite{FBB06,BBF06} we adopt in a first stage as the vector spin
variable the \textit{contravariant} components of the vector $\bar{S}_A^i$,
which are obtained by raising the index on $\bar{S}^A_i$ by means of the
spatial metric $\gamma_A^{ij}$, denoting the inverse of the covariant spatial
metric $\gamma^A_{ij}\equiv g^A_{ij}$ evaluated at point $A$ (i.e. such that
$\gamma_A^{ik}\gamma^A_{kj}=\delta^i_j$). Hence our initial spin variable is
\begin{equation}\label{eq:SA_covariant}
\bar{S}_A^i = \gamma_A^{ij}\bar{S}^A_j\, .
\end{equation}
This definition of the spin vector $\bar{\bm{S}}_A=(\bar{S}_A^i)$ agrees with
the choice already made in Refs.~\cite{OTO98, TOO01}.

At the leading SO approximation, the contravariant spin variables
$\bar{\bm{S}}_A$ defined by Eq.~\eqref{eq:SA_covariant} coincide with the spin
variables with constant magnitude broadly used in the literature (see, e.g.,
Ref.~\cite{K95}). At the next-to-leading order, the variables $\bar{\bm{S}}_A$
differ from constant magnitude spins and their relationship has been worked
out at 2PN order in Eq.~(7.4) of Ref.~\cite{BBF06}. In the present paper we
shall denote the constant magnitude spins by $\bm{S}_A$ (although they were
denoted $\bm{S}^\text{c}_A$ in Refs.~\cite{FBB06,BBF06}). We know that it is
actually better when presenting final results to switch to the constant
magnitude spins $\bm{S}_A$ since they have a simpler precession equation (and
turn out to be secularly conserved, i.e., over a radiation-reaction time
scale; see Ref.~\cite{W05} and Appendix \ref{sec:secular} below).

For two bodies ($A=1,2$) the relationship between the constant magnitude spins
and the original spin variables up to 1PN order is:
\begin{equation}\label{eq:Sbar}
\bm{S}_1 = \bar{\bm{S}}_1 + \frac{1}{c^2}\biggl[-\frac{1}{2}
  (v_1\bar{S}_1)\,\bm{v}_1 + \frac{G
    m_2}{r_{12}}\,\bar{\bm{S}}_1\biggr]
+\mathcal{O}\Bigl(\frac{1}{c^4}\Bigr)\,,
\end{equation}
together with the relation for the other particle obtained by exchanging all
particle labels. We denote by $\bm{v}_A=\ud\bm{y}_A/\ud t$ the coordinate
velocity of the particle $A$ (with mass $m_A$) and by
$r_{12}=\vert\bm{y}_1-\bm{y}_2\vert$ the relative distance. See
Ref.~\cite{BBF06} for more accurate formulas extending \eqref{eq:Sbar} to 2PN
order. In the present paper we shall consistently work only with the constant
magnitude spins $\bm{S}_A$.

In the case of binary systems it is convenient to pose
\begin{subequations}\label{eq:SSigma}
\begin{align} 
\bm{S} &= \bm{S}_1 + \bm{S}_2\,,\\
\bm{\Sigma} 
&= \frac{\bm{S}_2}{X_2} -
\frac{\bm{S}_1}{X_1}\,, 
\end{align}
\end{subequations}
where $X_1=m_1/m$ and $X_2=m_2/m$ (with $m=m_1+m_2$).  In addition we find it
useful to occasionally use the dimensionless spin variables
\begin{equation}\label{eq:ssigma}
\bm{s} = \frac{\bm{S}}{G\,m^2}\,, \qquad\bm{\sigma}
= \frac{\bm{\Sigma}}{G\,m^2}\,.
\end{equation}

\subsection{Multipole moments with spin-orbit effects}\label{sec:multspin}

The matter-source densities \eqref{eq:densities} depend on the
components of the stress-energy tensor. At the leading PN order, the spin
contribution therein (indicated by the subscript S) reduce to
\begin{subequations}\label{eq:sigmaN}\begin{align}
\mathop{\sigma}_\text{S} &=
-\frac{2}{c^3}\,\varepsilon_{ijk}\,v_1^i\,S_1^j\,\partial_k\delta_1
+1\leftrightarrow
2+\mathcal{O}\Bigl(\frac{1}{c^5}\Bigr)\,,\\ \mathop{\sigma}_\text{S}{}_{\!i}
&=
-\frac{1}{2c}\,\varepsilon_{ijk}\,S_1^j\,\partial_k\delta_1+1\leftrightarrow
2+\mathcal{O}\Bigl(\frac{1}{c^3}\Bigr)\,, \label{eq:sigmaiN}
\\ \mathop{\sigma}_\text{S}{}_{\!ij} &=
-\frac{1}{c}\,\varepsilon_{kl(i}\,v_1^{j)}\,S_1^k\,\partial_l\delta_1+
1\leftrightarrow 2+\mathcal{O}\Bigl(\frac{1}{c^3}\Bigr)\,,
\end{align}\end{subequations}
where $\delta_1(\bm{x},t)=\delta[\bm{x}-\bm{y}_1(t)]$ means the
three-dimensional Dirac delta-function evaluated on the particle 1, and
$1\leftrightarrow 2$ means the same quantity but corresponding to the particle
2.

In Ref.~\cite{BBF06} the SO terms have been computed in the
source mass quadrupole moment $I_{ij}$ up to next-to-leading 2.5PN order and
the source current quadrupole moment $J_{ij}$ up to next-to-leading 1.5PN
order. All the other source moments were computed at the leading SO
order. Those results are sufficient for our purpose. Actually, to compute the
specific contributions of tails we need only the moments at leading SO order,
given for general $\ell$ by
\begin{subequations}\label{eq:ILJLNexpl}
\begin{align}
\mathop{I}_\text{S}{}_{\!\!L} &= \frac{2\ell}{c^3(\ell+1)} \Bigl[\ell
v_1^i S_1^j \varepsilon_{ij\langle i_\ell}\,y_1^{L-1\rangle} \\ & -
(\ell-1) y_1^i S_1^j \varepsilon_{ij\langle
i_\ell} v_1^{i_{\ell-1}} y_1^{L-2\rangle} \Bigr] +
1\leftrightarrow 2 + \mathcal{O}\Bigl(\frac{1}{c^5}\Bigr) \, ,\nonumber\\
\mathop{J}_\text{S}{}_{\!\!L} &= \frac{\ell+1}{2c} y_1^{\langle
L-1} S_1^{i_\ell\rangle} + 1 \leftrightarrow 2 +
\mathcal{O}\Bigl(\frac{1}{c^3}\Bigr) \,.
\end{align}
\end{subequations}
Because the leading SO terms scale as $\mathcal{O}(1/c^3)$ in the mass source
moments, and as $\mathcal{O}(1/c)$ in the current source moments, the number
of non-linear terms needed in the radiative moments [Eqs.~\eqref{eq:deltaUV}
  below] is small. We refer to Sec.~V of \cite{BBF06} for higher-order
expressions of SO contributions of the source quadrupole moments.

\begin{figure}
\begin{center}
\includegraphics[width=\linewidth]{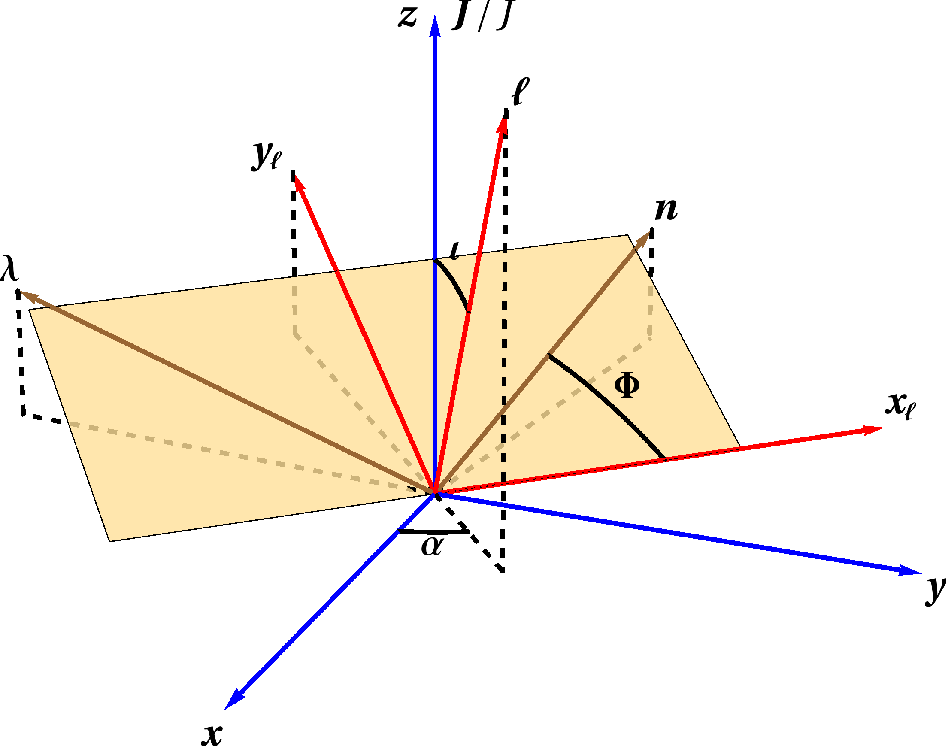}
\caption{We show (i) the source frame defined by the orthonormal basis
  $(\bm{x},\bm{y},\bm{z})$, (ii) the instantaneous orbital
  plane which is described by the orthonormal basis $(\bm{x}_\ell, \bm{y}_\ell,
  \bm{\ell})$, (iii) the moving triad $(\bm{n},\bm{\lambda},\bm{\ell})$, and 
  (iv) the direction of the total angular momentum $\bm{J}$ 
(agreeing by definition with the $z$--direction). Dashed lines show
  projections into the $x\mbox{--}y$ plane. 
  \label{figure:SourceFramePhase}}
\end{center}
\end{figure}

\subsection{Equations of motion with spin-orbit effects}\label{sec:eomspin}

Here we investigate the case where the binary's orbit is nearly circular,
i.e., whose radius is constant apart from small perturbations induced by the
spins (as usual we neglect the gravitational radiation damping at 2.5PN
order). We denote by $\bm{x}=\bm{y}_1-\bm{y}_2$ the relative position of the
particles (and $\bm{v}=\ud\bm{x}/\ud t$). Following Ref.~\cite{K95} we
introduce an orthonormal moving triad $\{\bm{n},\bm{\lambda},\bm{\ell}\}$
defined by $\bm{n}=\bm{x}/r$ as before,
$\bm{\ell}=\bm{L}_\mathrm{N}/\vert\bm{L}_\mathrm{N}\vert$ where
$\bm{L}_\mathrm{N}\equiv m \nu \,\bm{x}\times\bm{v}$ with $\nu = X_1 X_2$
denotes the Newtonian orbital angular momentum and $\nu$ the symmetric mass
ratio, and $\bm{\lambda}=\bm{\ell}\times\bm{n}$. Those vectors are represented
on Fig.~\ref{figure:SourceFramePhase}, which shows the geometry of the system.
The orbital frequency $\omega$ is defined for general, not necessarily
circular orbits, by $\bm{v}=\dot{r}\bm{n}+r\omega\bm{\lambda}$ where $\dot{r}$
represents the derivative of $r$ with respect to the coordinate time $t$. It
is also equal to the scalar product of $\bm{n}$ and $\bm{v}$ which we denote
as $(nv)=\dot{r}$. The components of the acceleration $\bm{a}=\ud\bm{v}/\ud t$
along the basis $\{\bm{n},\bm{\lambda},\bm{\ell}\}$ are then given by
\begin{subequations}\label{eq:acceleration_components}\begin{align}
\bm{n}\cdot\bm{a} &= \ddot{r} - r \omega^2\,,\\
\bm{\lambda}\cdot\bm{a} &= r \dot{\omega} + 2 \dot{r} \omega\,,\\
\bm{\ell}\cdot\bm{a} &= - r \omega \Bigl(\bm{\lambda}\cdot\frac{\ud
\bm{\ell}}{\ud t}\Bigr)\,.
\end{align}\end{subequations}
We project out the spins on this orthonormal basis, defining $\bm{S}= S_n
\bm{n} + S_\lambda \bm{\lambda} + S_\ell \bm{\ell}$ and similarly for
$\bm{\Sigma}$. Next we impose the restriction to quasi-circular precessing
orbits which is defined by the conditions 
$\ddot{r}=0=\dot{r}$ so that $v^2 = r^2\omega^2$ (neglecting radiation reaction
damping terms). In this way we find \cite{FBB06} that the equations of the
relative motion in the frame of the center-of-mass are
\begin{equation}\label{eq:a_circ}
\frac{\ud \bm{v}}{\ud t} = - \omega\,r\Bigl[\omega\,\bm{n} +
  \omega_\text{prec}\,\bm{\ell}\Bigr] + \mathcal{O}\Bigl(\frac{1}{c^6}\Bigr)
\, .
\end{equation}
There is no component of the acceleration along $\bm{\lambda}$. Comparing with
Eqs.~\eqref{eq:acceleration_components} in the case of circular orbits, we see
that $\omega$ is indeed the orbital frequency, while what we call the
``precessional frequency'' $\omega_\text{prec}=\bm{\lambda}\cdot \ud
\bm{\ell}/\ud t$ is proportional to the variation of $\bm{\ell}$ in the
direction of the velocity $\bm{v}=r\omega\bm{\lambda}$.  We know that
$\omega^2$ is given by
\begin{align}\label{eq:omega} 
\omega^2 &= \frac{G\,m}{r^3}\bigg\{1 + \gamma\left(-3+\nu\right) +
\gamma^{3/2}\left(-5 s_\ell-3 \delta \sigma_\ell\right)\biggr\}\nonumber\\&
+\mathcal{O}\Bigl(\frac{1}{c^4}\Bigr)\,,
\end{align}
where we denote $\delta \equiv X_1 - X_2 $ and $s_\ell\equiv (s\ell)
= \bm{s}\cdot\bm{\ell}$, where the spin variables are defined by
Eq.~\eqref{eq:ssigma}. The PN parameter is $\gamma\equiv G m/(r c^2)$ and we
have included only the 1PN non-spin term and the leading SO correction at
1.5PN order. On the other hand, we get \cite{FBB06}
\begin{equation}\label{eq:al} 
\omega_\text{prec} = - \omega\,\gamma^{3/2}\Bigl(7 s_n + 3 \delta\sigma_n\Bigr)+
\mathcal{O}\Bigl(\frac{1}{c^4}\Bigr)\,,
\end{equation}
where $s_n\equiv (s n) = \bm{s}\cdot\bm{n}$. At the leading 1.5PN SO order the
orbital frequency \eqref{eq:omega}, as well as $\omega_\text{prec}$, remain
unchanged if we were to substitute some other variables to the spins $\bm{S}$,
$\bm{\Sigma}$. However, when working at a higher PN approximation, it is more
convenient to use the spin variables $\bm{S}$, $\bm{\Sigma}$, built from the
constant magnitude spins. The main advantage of the spins $\bm{S}_A$ is that
they satisfy the usual-looking precession equations
\begin{equation}\label{eq:precession_equations}
\frac{\ud \bm{S}_A}{\ud t} = \bm{\Omega}_A\times
\bm{S}_A\,,
\end{equation}
showing that the spins precess around the direction of $\bm{\Omega}_A$, and at
the rate $\Omega_A=\vert\bm{\Omega}_A\vert$. The
equation \eqref{eq:precession_equations} could in principle be extended to any
PN order (at the linear SO level). The precession's angular-frequency vectors
$\bm{\Omega}_A$ have been computed up to the 2PN order for circular orbits in
Ref.~\cite{BBF06}. Here, we shall only need the 1PN leading order:
\begin{equation}\label{eq:Omega1_circ}
\bm{\Omega}_1 = \omega\,\gamma \,\biggl[ \frac{3}{4} +
  \frac{\nu}{2}  -\frac{3}{4} \delta
   \biggr]\,\bm{\ell} + \mathcal{O} \Bigl(\frac{1}{c^4}\Bigr) \,.
\end{equation}
To obtain $\bm{\Omega}_2$ we simply have to change $\delta$ into
$-\delta$. Both precession frequencies are constant in magnitude and
independent of the spins in the 1.5PN dynamics.

The equations of motion \eqref{eq:a_circ} and the precession equations
\eqref{eq:precession_equations} together leave invariant the total angular
momentum,
\begin{equation}\label{eq:J}
\bm{J}=\bm{L}+\frac{1}{c}\bm{S}\,,\qquad\frac{\ud\bm{J}}{\ud t}=0\, ,
\end{equation}
where $\bm{L}$ denotes the orbital angular momentum. For future reference we
give the components of $\bm{L}$ along the triad basis at 1PN order for
non-spin effects and at the leading 1.5PN order for spin ones
\cite{FBB06,BBF06}:
\begin{subequations}\label{eq:L}
\begin{align} \label{eq:Ll}
L_\ell &= \frac{G m^2 \nu}{c} x^{-1/2}\Bigl[1 +
  \Bigl(\frac{3}{2}+\frac{\nu}{6}\Bigr)x\Bigr]\,,\\ L_n &=
\frac{\nu\,x}{c}\Bigl[\frac{1}{2}S_n+\frac{1}{2}\delta \Sigma_n\Bigr]\,,\\
L_\lambda 
&= \frac{\nu\,x}{c}\Bigl[-{3}S_\lambda-\delta \Sigma_\lambda\Bigr]\,.
\end{align}\end{subequations}
Note that the components $L_n$ and $L_\lambda$ are due to spin effects arising
at order $\mathcal{O}(c^{-3})$. See Eq.~(7.10) of Ref.~\cite{BBF06}.

\section{Evolution of the triad $\{\bm{n}, \bm{\lambda}, \bm{\ell}\}$}
\label{sec:nlambdaell}

\begin{figure}
\begin{center}
\includegraphics[width=0.9\linewidth]{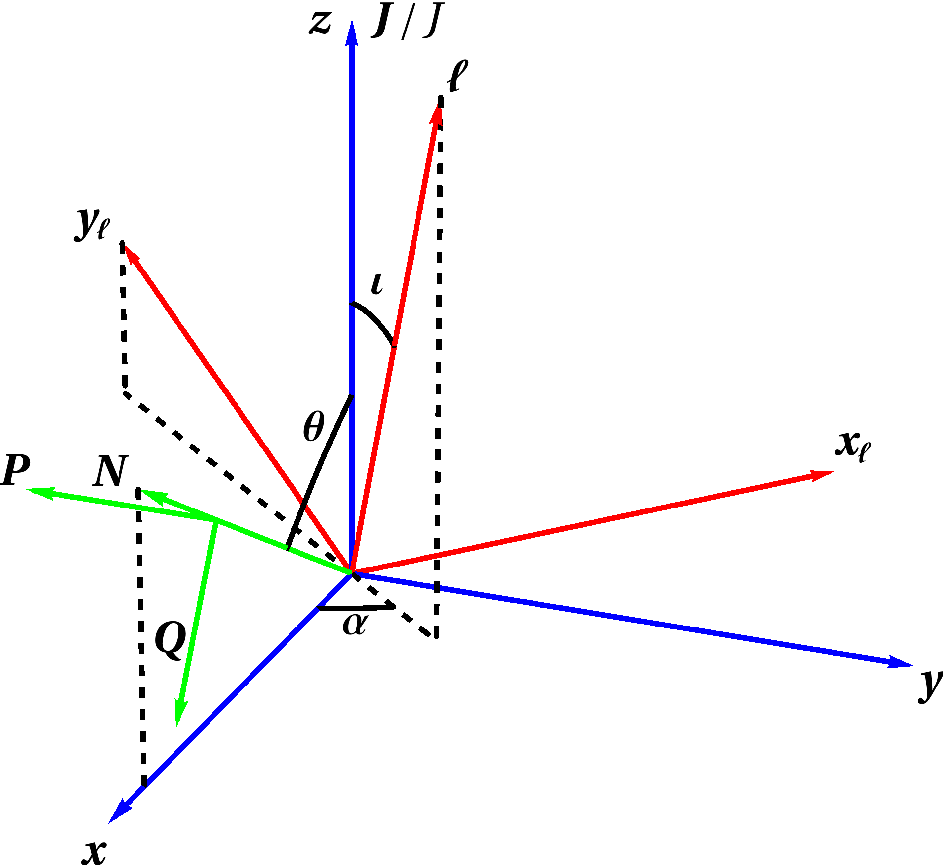}
\caption{ Similar as Fig.~\ref{figure:SourceFramePhase} but with the direction
  of the source $\bm{N}$ indicated together with a choice of convention for
  the two polarization vectors $\bm{P}$ and $\bm{Q}$.
  \label{figure:SourceFrame}}
\end{center}
\end{figure}
Using Eq.~\eqref{eq:a_circ} the time derivatives of the three moving triad
vectors $\{\bm{n}, \bm{\lambda}, \bm{\ell}\}$ can be expressed with respect to
that triad basis as
\begin{subequations} \label{eq:triad_evolution}
\begin{align}
\frac{\ud\bm{n}}{\ud t} &= \omega \,\bm{\lambda} \,
, \label{eq:triad_evolution_n} \\
\frac{\ud\bm{\lambda}}{\ud t} &= - \omega \,\bm{n} - \omega_\text{prec}
\,\bm{\ell} \, , \label{eq:triad_evolution_lambda} \\
\frac{\ud\bm{\ell}}{\ud t} &= \omega_\text{prec} \,\bm{\lambda} \, .
\end{align}
\end{subequations}
Equivalently, introducing the orbital precession vector
$\bm{\omega}=\omega\,\bm{\ell}$ and spin precession vector
$\bm{\omega}_\text{prec}=-\omega_\text{prec}\,\bm{n}$, these equations can be
written as
\begin{subequations} \label{eq:triad_evolution2}
\begin{align}
\frac{\ud\bm{n}}{\ud t} &= \bm{\omega}\times\bm{n} \, ,
\\ \frac{\ud\bm{\lambda}}{\ud t} &= \bigl(\bm{\omega}+
\bm{\omega}_\text{prec}\bigr)\times \bm{\lambda} \, ,
\\ \frac{\ud\bm{\ell}}{\ud t} &= \bm{\omega}_\text{prec} \times\bm{\ell} \, .
\end{align}
\end{subequations}
We recall that the spin precession frequency is given by Eq.~\eqref{eq:al} or
equivalently
\begin{equation}\label{prec}
\omega_\text{prec} = -  \omega \,x^{3/2}\Bigl(7 s_n + 3 \delta \sigma_n\Bigr)+
\mathcal{O}\Bigl(\frac{1}{c^4}\Bigr)\, ,
\end{equation}
where we pose $x\equiv(G m \omega/c^3)^{2/3}$ which defines a gauge-invariant PN
parameter agreeing with $\gamma$ at leading order.

We shall now solve the evolution equations for the moving triad $\{\bm{n},
\bm{\lambda}, \bm{\ell}\}$ at the 1.5PN order in the conservative dynamics
(i.e., neglecting radiation reaction). We recall that the spin variables
we use in this calculation are those with constant magnitude, denoted here
$\bm{S}_A$.

It is convenient to introduce a fixed (inertial) orthonormal basis $\{ \bm{x},
\bm{y}, \bm{z} \}$; see Figs.~\ref{figure:SourceFramePhase}
and~\ref{figure:SourceFrame}. For a given value of the total angular momentum
$\bm{J}$ (a constant vector), and of the direction $\bm{N}=\bm{X}/R$ of the
detector as seen from the source, a canonical choice of the basis vectors is
as follows: (i) $\bm{z}$ is defined to be the normalized value of $\bm{J}$,
namely $\bm{J}/J$; (ii) $\bm{y}$ is orthogonal to the plane spanned by
$\bm{N}$ and $\bm{z}$ and points to the direction that corresponds to the
positive orientation of the acute angle $\langle \bm{z}, \bm{N} \rangle$, i.e.
$\bm{y} = \bm{z} \times \bm{N}/|\bm{z} \times \bm{N}| $; (iii) $\bm{x}$
completes the triad. We see that $\bm{x}$, $\bm{z}$ and $\bm{N}$ are coplanar
by construction. Then, we introduce the standard spherical coordinates with
the inclination angle measured from the zenith direction $\bm{z}$ and the
azimuthal angle measured from $\bm{x}$. The spherical coordinates of $\bm{N}$
and $\bm{\ell}$ are denoted as $(\theta, \varphi)$ and $(\alpha, \iota)$
respectively, and since $\bm{N}$ lies in the same plane as $\bm{x}$ and
$\bm{z}$, we have $\varphi=0$ (see Fig.~\ref{figure:SourceFrame}). Since
$\iota$ is the angle between the total and orbital angular momenta, we have
\begin{equation} \label{eq:sin_iota}
\sin \iota = \frac{\vert\bm{J} \times \bm{\ell}\vert}{J} \, .
\end{equation}
The angles $(\alpha, \iota)$ are referred to as the precession angles. 

We now derive the time evolution of our triad vectors from that of the
precession angles $(\alpha, \iota)$, and of an appropriate phase $\Phi$ that
specifies the position of $\bm{n}$ with respect to some reference
direction. Following Ref.~\cite{ABFO08}, we introduce the unit vectors
\begin{equation}\label{xell}
\bm{x}_\ell = \frac{\bm{J} \times \bm{\ell}}{\vert \bm{J} \times \bm{\ell}
  \vert}\, , \qquad \bm{y}_\ell = \bm{\ell} \times \bm{x}_\ell \, ,
\end{equation}
such that $\{\bm{x}_\ell, \bm{y}_\ell, \bm{\ell}\}$ is an orthonormal basis. 
The phase angle $\Phi$ is defined by (see Fig.~\ref{figure:SourceFramePhase}):
\begin{equation}
\Phi = \langle\bm{x}_\ell , \bm{n}\rangle = \langle\bm{y}_\ell ,
\bm{\lambda}\rangle \, .
\end{equation}
The rotation takes place in the instantaneous orbital plane spanned by
$\bm{n}$ and $\bm{\lambda}$,
and we have
\begin{subequations} \label{eq:n_lambda}
\begin{align}
&\bm{n} = \cos \Phi \, \bm{x}_\ell + \sin \Phi \, \bm{y}_\ell\, , \label{eq:n}
  \\ &\bm{\lambda} = -\sin \Phi \, \bm{x}_\ell + \cos \Phi \, \bm{y}_\ell
  \,,\label{eq:lambda}
\end{align}
\end{subequations}
from which we deduce
\begin{equation} \label{eq:exp_Phi}
e^{-\ui \,\Phi} = \bm{x}_\ell\cdot\bigl(\bm{n}+\ui \bm{\lambda}\bigr) =
\frac{J_\lambda - \ui\,J_n }{\sqrt{J_n^2 + J_\lambda^2}}\,.
\end{equation}
Combining \eqref{eq:exp_Phi} with \eqref{eq:sin_iota} we also get
\begin{equation} \label{eq:sin_iota_exp_Phi}
\sin\iota\,e^{-\ui \,\Phi} = \frac{J_\lambda - \ui\,J_n }{J}\,.
\end{equation}

By identifying the right-hand sides of Eqs.~\eqref{eq:triad_evolution} or
\eqref{eq:triad_evolution2} with the time-derivatives of the
identities~\eqref{eq:n_lambda} we obtain the following system of equations for
the variations of $\alpha$, $\iota$ and $\Phi$, equivalent to the
system~\eqref{eq:triad_evolution},
\begin{subequations} 
\begin{align}
\label{alphadot}
\frac{\ud \alpha}{\ud t} &= -\omega_{\rm prec}\,\frac{\sin \Phi}{\sin \iota}\,,\\
\label{iotadot}
\frac{\ud \iota}{\ud t} &= - \omega_{\rm prec}\, \cos \Phi\,,\\
\label{Phidot}
\frac{\ud \Phi}{\ud t} &= \omega + \omega_{\rm prec}\,\frac{\sin \Phi}{\tan
  \iota}\,.
\end{align}
\end{subequations}

On the other hand, using the total angular momentum \eqref{eq:J} together with
the components of the orbital angular momentum given by Eqs.~\eqref{eq:L} ---
notably the fact that $L_n$ and $L_\lambda$ are due to SO terms dominantly of
order $\mathcal{O}(c^{-3})$, we deduce that $\sin\iota$ is a small quantity of
order $\mathcal{O}(1/c)$. From this fact, we conclude by direct integration of
the sum of Eq.~\eqref{alphadot} and Eq.~\eqref{Phidot} that
\begin{equation}\label{Phi}
\Phi + \alpha = \phi +
\mathcal{O}\Bigl(\frac{1}{c^4}\Bigr) \, ,
\end{equation}
in which we have defined the ``carrier'' phase as 
\begin{equation}\label{carrier}
\phi = \int \omega \, \ud t = \omega (t-t_0) + \phi_0 \, ,
\end{equation}
with $\phi_0$ the value of the carrier phase at some arbitrary initial time
$t_0$. We recall that the orbital frequency \eqref{eq:omega} is constant in
first approximation for circular motion.

The combination $\Phi+\alpha$ being known by Eq.~\eqref{Phi}, we can further
express the precession angles $\iota$ and $\alpha$ in first approximation in
terms of the components $S_n$ and $S_\lambda$ of the total spin
$\bm{S}=\bm{S}_1+\bm{S}_2$. From \eqref{eq:sin_iota} we find (discarding
non-linear spin contributions)
\begin{equation} \label{siniota}
\sin\iota = \frac{\sqrt{S_n^2 + S_\lambda^2}}{c L_\text{N}} +
\mathcal{O}\Bigl(\frac{1}{c^3}\Bigr)\,,
\end{equation}
where we recall that $L_\text{N}=m\nu r^2\omega$ denotes the Newtonian orbital
angular momentum. On the other hand, using also
Eq.~\eqref{eq:sin_iota_exp_Phi} and the relation \eqref{Phi} we obtain
at leading order
\begin{equation} \label{alpha}
e^{\ui \,\alpha} = \frac{S_\lambda - \ui\,S_n
}{\sqrt{S_n^2+S_\lambda^2}}\,e^{\ui \,\phi} +
\mathcal{O}\Bigl(\frac{1}{c^2}\Bigr)\,.
\end{equation}
[See also the more precise 
equations~\eqref{eq:sin_iota_exp_Phi_15PN}--\eqref{eq:Jplus}.]

It remains now to obtain the explicit time variation of the components of the
individual spins $S_n^A$, $S_\lambda^A$ and $S_\ell^A$. Using \eqref{siniota}
and \eqref{alpha} [and also \eqref{Phi}] we shall then be able to obtain the
explicit time variation of the precession angles and phase. Combining
\eqref{eq:precession_equations} and~\eqref{eq:triad_evolution} we obtain
the precession equations for the three unknowns $S^A_n$, $S^A_\lambda$ and
$S^A_\ell$ in the form of the following first-order system (valid at any PN
approximation)
\begin{subequations}\label{spinevolv}
\begin{align}
\frac{\ud S_n^A}{\ud t} &= \bigl(\omega - \Omega_A \bigr)
S_\lambda^A\,,\\ \frac{\ud S_\lambda^A}{\ud t} &= - \bigl(\omega - \Omega_A
\bigr) S_n^A - \omega_\text{prec}\,S_\ell^A\,,\\ \frac{\ud S_\ell^A}{\ud t} &=
\omega_\text{prec}\,S_\lambda^A\,,
\end{align}
\end{subequations}
where $\Omega_A$ is the norm of the precession vector of the spin $A$ as given
by \eqref{eq:Omega1_circ}, and the precession frequency $\omega_\text{prec}$
is explicitly given by \eqref{prec}. Actually the terms involving
$\omega_\text{prec}$ in the right-hand sides of \eqref{spinevolv} can be
neglected because they are quadratic in the spins. Thus, staying at the linear
SO level, we find that the equations \eqref{spinevolv} can be decoupled and
integrated as
\begin{subequations}\label{spinperp}
\begin{align}
S^A_n &= S_\perp^A \cos\psi_A \, ,\\
S^A_\lambda &= - S_\perp^A \sin\psi_A \, ,\\
S^A_\ell &= S_{\parallel}^A \, .
\end{align}\end{subequations}
Here $S_\perp^A$ and $S_{\parallel}^A$ denote two constants for each spins $A$,
and agree with the projections (which are constant at the linear SO level) of
the spins perpendicular and parallel to the direction of the orbital angular
momentum $\bm{\ell}$. The phase of each of the spins is given by
\begin{equation}\label{psiA}
\psi_A = (\omega - \Omega_A)(t-t_0)+\psi^0_A\,,
\end{equation}
where $\psi^0_A$ is the constant initial phase at time $t_0$.

With those results we obtain an explicit solution for the precession angles by
substituting Eqs.~\eqref{spinperp} into the results \eqref{siniota} and
\eqref{alpha}. We find that $\iota(t)$ is given at the 0.5PN level by
\begin{align}\label{iotasol}
\sin\iota &=
\frac{x^{1/2}}{\nu}\sqrt{(s_\perp^1)^2+(s_\perp^2)^2+2s_\perp^1s_\perp^2
  \cos(\psi_1-\psi_2)}\nonumber\\ &+ \mathcal{O}\Bigl(\frac{1}{c^3}\Bigr)\,,
\end{align}
where we recall that $s_\perp^A=S_\perp^A/(Gm^2)$. Knowing $\iota(t)$
we deduce $\alpha(t)$ from
\begin{equation}\label{alphasol}
\sin\iota \,e^{\ui \alpha} = -\ui \frac{x^{1/2}}{\nu} e^{\ui
  \phi}\,\Bigl(s_\perp^1 e^{-\ui \psi_1} +s_\perp^2 e^{-\ui \psi_2}\Bigr) +
\mathcal{O}\Bigl(\frac{1}{c^3}\Bigr)\,.
\end{equation}
The difference of spin phases $\psi_{12}\equiv\psi_1-\psi_2$ readily follows
from Eq.~\eqref{psiA} and Eq.~\eqref{eq:Omega1_circ} at 1PN order as
\begin{equation}\label{psi12}
\psi_{12} = \psi^0_{12} + \frac{3}{2}\omega\,x\,\delta (t-t_0) +
\mathcal{O}\Bigl(\frac{1}{c^4}\Bigr)\,.
\end{equation}
Moreover, Eq.~(\ref{eq:sin_iota_exp_Phi}) can be written more explicitly
  at the 1.5PN level as
\begin{equation} \label{eq:sin_iota_exp_Phi_15PN}
\sin \iota \, e^{-\ui \,\Phi} = -\ui
\frac{J_+}{L_\ell} + \mathcal{O} \Bigl(\frac{1}{c^4}\Bigr)\,,
\end{equation}
where $J_+ \equiv J_n + \ui \, J_\lambda$ is given at the 1.5PN order by
\begin{align} \label{eq:Jplus}
& J_+  =
\frac{S^1_{\perp}}{c} \bigg\{ e^{-\ui \psi_1} \bigg[ 1 + x
  \Bigl(-\frac{1}{8} - \frac{3}{4} \nu + \frac{\delta}{8} \Bigr) \bigg] \\ &
\qquad + e^{\ui \psi_1 } x \Bigl(\frac{3}{8}+\frac{\nu}{4} - \frac{3\delta}{8}
\Bigr) \bigg\} + 1 \leftrightarrow 2 + \mathcal{O}\Bigl(\frac{1}{c^4}\Bigr)\,,
\nonumber \end{align}
and the 1PN orbital angular momentum $L_\ell$ is known from Eq.~\eqref{eq:Ll}.

As a check of the previous solution we observe that if we take the time
derivative of Eq.~\eqref{eq:sin_iota}, then evaluate the total angular
momentum $\bm{J}$ given by \eqref{eq:J} together with the components of the
orbital angular momentum $\bm{L}$ provided in \eqref{eq:L}, and use the
solution \eqref{spinevolv} for the evolution of the spin components, we obtain
\begin{equation}
\frac{\ud \iota}{\ud t} = - \omega_{\rm prec}\,\frac{S_\lambda}{\sqrt{S_n^2 +
    S_\lambda^2}} + \mathcal{O} \Bigl(\frac{1}{c^4}\Bigr)\,,
\end{equation}
which is consistent with \eqref{iotadot} once \eqref{eq:exp_Phi} is employed. 

Finally we express the triad vectors $\bm{n}(t)$, $\bm{\lambda}(t)$ and
$\bm{\ell}(t)$ in terms of the precession angles and the carrier phase, and in
terms of the initial triad and angles at the initial instant $t_0$, modulo
terms of order $\mathcal{O}(c^{-4})$. To do this we notice that the triad
$\{\bm{n}, \bm{\lambda},\bm{\ell}\}$ at time $t$ is obtained from the inertial
triad $\{\bm{x}, \bm{y}, \bm{z}\}$ by the rotation associated with the three
Euler angles $\alpha$, $\iota$ and $\Phi$. Similarly the initial triad
$\{\bm{n}_0, \bm{\lambda}_0,\bm{\ell}_0\}$ at time $t_0$ is obtained by the
rotation associated with $\alpha_0$, $\iota_0$ and $\Phi_0$. So, combining
those two rotations we readily obtain $\{\bm{n}, \bm{\lambda},\bm{\ell}\}$ in
terms of $\{\bm{n}_0, \bm{\lambda}_0,\bm{\ell}_0\}$. Using Eq.~\eqref{Phi} to
eliminate the phase $\Phi$ in favor of the carrier phase $\phi$ --- this
introduces small remainder terms $\mathcal{O}(c^{-4})$ --- and neglecting all
terms quadratic in the spins, we get
\begin{widetext}
\begin{subequations} \label{eq:n_lambda_l}
\begin{align}
&\bm{n} = \cos (\phi - \phi_0) \bm{n}_0 + \sin (\phi - \phi_0) \bm{\lambda}_0
  + \Bigl( \sin \iota \sin (\phi-\alpha) - \sin \iota_0 \sin
  (\phi-\alpha_0)\Bigr) \bm{\ell}_0 + \mathcal{O} \Bigl(\frac{1}{c^4} \Bigr)\,
  , \\ &\bm{\lambda} = - \sin (\phi - \phi_0) \bm{n}_0 + \cos (\phi - \phi_0)
  \bm{\lambda}_0 + \Bigl(\sin \iota \cos (\phi - \alpha) -\sin \iota_0 \cos
  (\phi - \alpha_0) \Bigr) \bm{\ell}_0 + \mathcal{O} \Bigl(\frac{1}{c^4}
  \Bigr)\, ,\\ &\bm{\ell} = \bm{\ell}_0 + \Bigl(-\sin \iota \sin
  (\phi_0-\alpha) + \sin \iota_0 \sin (\phi_0 - \alpha_0)\Bigr) \bm{n}_0 +
  \Bigl(-\sin \iota \cos (\phi_0 - \alpha) + \sin \iota_0 \cos
  (\phi_0-\alpha_0)\Bigr) \bm{\lambda}_0 + \mathcal{O} \Bigl(\frac{1}{c^4}
  \Bigr)\,.
\end{align}
\end{subequations}
\end{widetext}
Since we have found in Eqs.~\eqref{iotasol}--\eqref{alphasol} an explicit
solution for the precession angles $\iota(t)$ and $\alpha(t)$, the time
dependence of $\bm{n}$, $\bm{\lambda}$ and $\bm{\ell}$ is completely
known. Note that in practical computations it is often more convenient to work
not with $\bm{n}$ and $\bm{\lambda}$ but with the complex null vector
$\bm{m}=(\bm{n}+\ui\,\bm{\lambda})/\sqrt{2}$ and its conjugate.

\section{Computation of the waveform}
\label{sec:comput}

Here we shall compute the SO terms coming from all non-linear (i.e. of formal
order $G^2$) contributions associated with tails consistent with the 2.5PN and
3PN orders in the waveform. We shall need only to focus on the tails entering
the mass and current quadrupoles $U_{ij}$ and $V_{ij}$ (having $\ell=2$) and
on the current octupole $V_{ijk}$ ($\ell=3$). They indeed contain, when
specialized to spinning compact binary systems, the SO contributions we are
interested in. The reason is that the leading SO terms start at the 0.5PN
order $\mathcal{O}(1/c)$ in the current moments, but only at the 1.5PN order
$\mathcal{O}(1/c^3)$ in the mass moments [see Eqs.~\eqref{eq:ILJLNexpl}
  above].

In addition to the tail integrals shown in Eq.~\eqref{eq:tails}, we shall also
compute some terms of order $G^2$ at 2.5PN or 1.5PN order, but which are
non-hereditary, i.e., merely depend on the instantaneous retarded time $T_R$.
Those corrections are given in full form by Eqs.~(5.4)--(5.6) of
Ref.~\cite{BFIS08}, but here we shall need only, for the same reason as
before, two terms involving the source current dipole moment or angular
momentum $J_i$; the other terms will not contribute to the SO effect at 3PN
order. Furthermore, we have to include some additional corrections of similar
nature originating from the differences between the canonical and the source
moments in Eqs.~\eqref{eq:MLSL}.  Those are given in the general case by
Eqs.~(5.9)--(5.10) of Ref.~\cite{BFIS08}, but we can check that only the
contribution in the mass quadrupole $U_{ij}$ \textit{a priori} matters here.

The relevant contributions in the radiative moments, including only the terms
needed for the applications below (see Ref.~\cite{BFIS08} for more complete
expressions) read:
\begin{widetext}
\begin{subequations} \label{eq:deltaUV}
\begin{align}
\delta U_{ij} &= I^{(2)}_{ij} + \frac{2 G m}{c^3} \int_{-\infty}^{T_R} \ud t
\biggl[\ln \biggl(\frac{T_R-t}{2\tau_0}\biggr) + \frac{11}{12} \biggr]
I^{(4)}_{ij}(t) + \frac{G}{c^5} \biggl( \frac{1}{3} \varepsilon_{ab\langle
  i} I^{(4)}_{j \rangle a} J_b + 4 \left[W^{(2)} I_{ij} - W^{(1)}
  I_{ij}^{(1)}\right]^{(2)} \biggr)\, , \label{eq:deltaUij} \\ \delta V_{ij}
&= J^{(2)}_{ij} + \frac{2 G m}{c^3} \int_{-\infty}^{T_R} \ud t \biggl[\ln
  \biggl(\frac{T_R-t}{2\tau_0}\biggr) + \frac{7}{6} \biggr]
J^{(4)}_{ij}(t) \,,\label{eq:deltaVij} \\ \delta V_{ijk} &= J^{(3)}_{ijk} +
\frac{2 G m}{c^3} \int_{-\infty}^{T_R} \ud t \biggl[\ln
  \biggl(\frac{T_R-t}{2\tau_0}\biggr) + \frac{5}{3} \biggr]
J^{(5)}_{ijk}(t) - \frac{2 G }{c^3} J_{\langle i} I^{(4)}_{jk \rangle}\,
, \label{eq:deltaVijk}
\end{align}
\end{subequations}
\end{widetext}
where we have replaced $M$ with $m=m_1+m_2$ which is valid at the dominant
order. Moreover, we shall find that the terms in the right-hand side of
Eq.~\eqref{eq:deltaUij} which depend on the moment $W$ vanish at the
considered order. Indeed, by inserting the value of $\sigma_i$ from
\eqref{eq:sigmaiN} into the potential $W$ defined by \eqref{eq:W} and
integrating by part, we find that the result is zero.

The corresponding gravitational-waveform, for which all three moments
$U_{ij}$, $V_{ij}$ and $V_{ijk}$ are important, is then given by
\begin{align} \label{eq:hTT}
\delta h^\mathrm{TT}_{ij} &= \frac{2G}{c^4 R} \mathcal{P}^\mathrm{TT}_{ijkl} \biggl[
  \delta U_{kl} - \frac{4}{3c} N_a \varepsilon_{ab \langle k} \, \delta V_{l \rangle
    b} \nonumber \\&\qquad - \frac{1}{2c^2} N_{am} \,\varepsilon_{ab
    \langle k} \, \delta V_{l \rangle bm} \biggr] \,.
\end{align}
The main task consists in evaluating the tail integrals \eqref{tailint} whose
Fourier transforms have already been obtained in Eq.~\eqref{tailsplitres}. The
result \eqref{tailsplitres} heavily relied on a physical assumption concerning
the system in the remote past, namely that it was formed by freely falling
incoming particles, see Eq.~\eqref{Ipast}.

We reviewed in Sec.~\ref{sec:tailint} that one can insert in the result
\eqref{tailsplitres} the binary's \textit{current} frequency spectrum, i.e. at
time $T_R$, modulo small error terms of the order of the adiabatic parameter
of the inspiral, or, more precisely, of negligible order $\mathcal{O}(\ln
c/c^5)$. In the case of spinning compact binaries this means that we have to
include in the spectrum the current orbital frequency
$\omega\equiv\omega(T_R)$, and also the precession frequencies
$\Omega_1\equiv\Omega_1(T_R)$ and $\Omega_2\equiv\Omega_2(T_R)$ of the two
spins. This follows from the explicit solution of the triad
$\{\bm{n},\bm{\lambda},\bm{\ell}\}$ and of the precession equations (see
Sec.~\ref{sec:appl}).[Notice that the precession angles $\alpha$ and $\iota$
always appear through the product $\sin \iota \, e^{\ui \alpha}$ given by
equations such as~\eqref{alphasol}.] Hence we can take for the
Fourier components of the multipole moments
\begin{equation}
\widetilde{I}_{L}(\Omega)=2\pi\sum_{n,n_1,n_2}A_{L}^{n,n_1,n_2}\,
\delta(\Omega-\omega_{n,n_1,n_2})\,,
\end{equation}
where the frequency modes are some
$\omega_{n,n_1,n_2}=n\,\omega+n_1\,\Omega_1+n_2\,\Omega_2$. The result
\eqref{tailsplitres} then becomes
\begin{align}\label{taildecomp}
&\mathcal{U}_{L}(T_R) = \sum_{n,n_1,n_2}\ui\,
  A_{L}^{n,n_1,n_2}(-\ui\omega_{n,n_1,n_2})^{\ell+1}\,e^{-\ui\,\omega_{n,n_1,n_2}T_R}
  \nonumber\\&\times\left[\frac{\pi}{2} \text{s}(\omega_{n,n_1,n_2}) + \ui
    \Bigl(\ln(2\vert\omega_{n,n_1,n_2}\vert\tau'_0)+\gamma_\text{E}\Bigr)\right]\,.
\end{align}
We recall from Eq.~\eqref{eq:Omega1_circ} that the precession frequencies
$\Omega_1$ and $\Omega_2$ are small quantities of order 1PN. This means in
particular that because of the explicit factor $\omega^{\ell+1}_{n,n_1,n_2}$
in Eq.~\eqref{taildecomp} (which arises from taking the time derivatives of
the multipole moment and integrating), the modes for which $n=0$ in tail
integrals are very small, at least of order 4.5PN, and can be neglected.

The SO terms in the radiative tails originate primarily from the spins present
in the sources of the integrals, i.e. appropriate derivatives of multipole
moments as shown in Eqs.~\eqref{eq:deltaUV}. The SO contributions in the
source moments have been given in Eqs.~\eqref{eq:ILJLNexpl}. However there are
other crucial SO terms which originate from the non-spin parts of the moments.
They come from time differentiations of the triad vectors using the evolution
equations~\eqref{eq:triad_evolution}, which produce spin terms contained in
the precession frequency $\omega_\text{prec}$, and from time differentiations
of spins themselves \textit{via} the precession equations
\eqref{eq:precession_equations}. SO contributions may also be generated by the
tail integration itself due to the precession of the triad basis
$\{\mathbf{n},\bm{\lambda},\bm{\ell}\}$ according to the formula
\eqref{eq:n_lambda_l}. On the other hand, other SO terms come from the
reduction to circular orbits when we eliminate the orbital separation $r$ in
favor of a function of the orbital frequency $\omega$ obtained from inverting
the relation \eqref{eq:omega}. Note that the two latter corrections being of
1.5PN relative order, they cannot come from anywhere but the tail integral of
the Newtonian quadrupole moment.

During the practical computation, we make explicit the time dependence of the
derivatives of multipole moments, computed in the center-of-mass frame as
functions of the relative position, the relative velocity and both spins. For
circular orbits, $\bm{x}$ and $\bm{v}$ depend only on $r$ and $\omega$, which
are approximately constant on dynamical time-scales, and on the unit vectors
$\bm{n}$ and $\bm{\lambda}$. Thus, the whole time dependence arises through
that of $\bm{n}$ and $\bm{\lambda}$ (and
$\bm{\ell}=\bm{n}\times\bm{\lambda}$), and is provided by our explicit
solution \eqref{eq:n_lambda_l}, together with the precessing angles
$\alpha(t)$ and $\iota(t)$ given by Eqs.~\eqref{iotasol}--\eqref{alphasol}, 
or~\eqref{eq:sin_iota_exp_Phi_15PN} and~\eqref{eq:Jplus} with more precision.

\begin{widetext}
The complete results for the spin dependent parts of the radiative moments, in
which we use the short-hand notation for spins \eqref{eq:ssigma} and where the
basis vectors $\{\bm{n},\bm{\lambda},\bm{\ell}\}$ are evaluated at the current
time $T_R$, are then
\begin{subequations}\label{eq:UVspintail}
\begin{align}
  \delta U_{ij} &= 2 m \nu x^4 c^2 \bigg[ \frac{1}{3} (71 s_n + 35 \delta
  \sigma_n) 
    \bigg(\pi \,n^{(i} \ell^{j)} -2 \lambda^{(i} \ell^{j)} \Bigl(\ln(4\omega
    \tau_0) + \gamma_\text{E} - \frac{11}{12} \Bigr) \bigg) \nonumber \\ &
    \qquad \qquad - \frac{1}{3} (29 s_\lambda + 17 \delta \sigma_\lambda)
    \bigg(\pi \,\lambda^{(i} \ell^{j)} + 2 n^{(i} \ell^{j)} \Bigl(\ln(4\omega
    \tau_0) + \gamma_\text{E} - \frac{11}{12}\Bigr) \bigg) \nonumber \\ &
    \qquad \qquad + 4 \Bigl(s_\ell + \frac{\delta}{3} \sigma_\ell \Bigr)
    \bigg(-4 n^{(i} \lambda^{j)} \Bigl(\ln(4 \omega \tau_0) + \gamma_\text{E}
    - \frac{11}{12} \Bigr) + \pi (n^{ij} - \lambda^{ij})\bigg) \nonumber \\ &
    \qquad \qquad -2 \bigg(n^{(i} \ell^{j)} \Bigl(\frac{19}{3} s_\lambda + 3
    \delta \sigma_\lambda \Bigr) + \lambda^{(i} \ell^{j)} \Bigl(\frac{19}{3}
    s_n + 3 \delta \sigma_n \Bigr) \bigg) - \frac{8}{3} s_\ell \,n^{(i}
    \lambda^{j)} \bigg] \, , \\ \delta V_{ij} &= - 3 m \nu x^{7/2} c^3
  \bigg[\lambda^{\langle i} \sigma^{j\rangle} \Bigl( \ln (2\omega \tau_0) +
  \gamma_\text{E} -
    \frac{7}{6} \Bigr) - \frac{\pi}{2} n^{\langle i} \sigma^{j \rangle} \bigg]
  \, , \\ \delta V_{ijk} &= - 16 m \nu x^4 c^4 \bigg[s^{\langle k} (n^{ij \rangle} -
    \lambda^{ij \rangle})+ 2 (s^{\langle k} + \delta \sigma^{\langle k})
    \bigg( (n^{ij \rangle} - \lambda^{ij \rangle}) \Bigl(\ln(4 \omega \tau_0)
    + \gamma_\text{E} - \frac{5}{3}\Bigr) + \pi \,n^{i} \lambda^{j \rangle}
    \bigg) \bigg] \, .
\end{align}\end{subequations}
Insertion of the above quantities into Eq.~\eqref{eq:hTT} yields the
non-linearly induced SO contributions at 2.5PN and 3PN orders in the waveform
as
\begin{align}\label{eq:hTTspintail}
\delta\hat{h}_{ij}^\text{TT} &= \Biggl\{x^{5/2} \biggl[\lambda^k \Bigl(\ln (2
  \tau_0 \omega) + \gamma_\text{E} - \frac{7}{6} \Bigr) - \frac{\pi}{2} n^k
  \biggr] \biggl[(\bm{N} \times \bm{\sigma})^i \delta_{jk} + N^a
  \varepsilon_{aki} \sigma^j \biggr] \nonumber \\ &\quad + x^3
\bigg[\frac{4}{3} \bigg(s^k (n^{cd} - \lambda^{cd}) + 2 (s^k + \delta
  \sigma^k) \Bigl( (n^{cd} - \lambda^{cd}) \Bigl(\ln (4\tau_0 \omega) +
  \gamma_\text{E} - \frac{5}{3} \Bigr) + \pi n^{(c} \lambda^{d)} \Bigr) \bigg)
  \times \nonumber \\ & \qquad \qquad \times \Bigl(N^{ak} \varepsilon_{aci}
  \delta_{jd} + N^{ca} \varepsilon_{aki} \delta_{jd} + N^{ad}
  \varepsilon_{aci} \delta_{jk} - \frac{2}{5} N^{aj} \varepsilon_{iac}
  \delta_{kd} \Bigr) \nonumber \\ & \qquad \qquad + 4 \Bigl(s_\ell +
  \frac{\delta \sigma_\ell}{3} \Bigr) \Bigl( -4 n^i \lambda^j \Bigl(\ln (4
  \tau_0 \omega) + \gamma_\text{E} - \frac{11}{12} \Bigr) + \pi (n^{ij} -
  \lambda^{ij} ) \Bigr) - \frac{8}{3} s_\ell n^i \lambda^j \nonumber \\ &
  \qquad \qquad -2 \Bigl( n^i \ell^j \Bigl( \frac{19}{3} s_\lambda + 3
  \delta\, \sigma_\lambda \Bigr) + \lambda^i \ell^j \Bigl( \frac{19}{3} s_n +
  3 \delta \sigma_n \Bigr) \Bigr) \nonumber \\ & \qquad \qquad + \frac{1}{3}
  (71 s_n + 35 \delta \sigma_n) \Bigl(\pi n^i \ell^j - 2 \lambda^i \ell^j
  \Bigl( \ln (4 \omega \tau_0) + \gamma_\text{E} - \frac{11}{12} \Bigr) \Bigr)
  \nonumber \\ & \qquad \qquad - \frac{1}{3} (29 s_\lambda + 17 \delta
  \sigma_\lambda) \Bigl(\pi \lambda^i \ell^j + 2 n^i \ell^j \Bigl( \ln (4
  \omega \tau_0) + \gamma_\text{E} - \frac{11}{12} \Bigr)
  \Bigr)\bigg]\Biggr\}^\text{TT}\,,
\end{align}
\end{widetext}
for which we have conveniently introduced the rescaled waveform $\delta
\hat{h}_{ij}^\text{TT}$ defined by
\begin{align} \label{eq:hTTspintail_true}
\delta h_{ij}^\text{TT} = \frac{4 G m \nu x}{R c^2}
\,\delta\hat{h}_{ij}^\text{TT} \, .
\end{align}
The two gravitational-wave polarizations $h_+$ and $h_\times$ are given in
Appendix \ref{sec:polarizations}. We have checked that the test-particle limit
$\nu \to 0$ of the $-2$ spin-weighted spherical modes resulting from the above
waveform agrees with the results of Ref.~\cite{TSTS96} (given explicitly in
Ref.~\cite{Pan:2010hz}) based on black-hole perturbation theory.

\section{Energy flux and orbital phasing}
\label{sec:Eflux}

The case of the gravitational energy flux is simpler than for the waveform,
notably because we need only the contributions from the mass and current
quadrupole moments, i.e.
\begin{align} \label{eq:flux_corrections}
\delta\mathcal{F} = \frac{G}{c^5} \biggl[\frac{2}{5}  U^{(1)}_{ij}
  \delta U^{(1)}_{ij} + \frac{32}{45 c^2}  V^{(1)}_{ij} \delta
  V^{(1)}_{ij} \biggr] \, .
\end{align}
The 3PN SO effects in the energy flux have been computed in two different
ways. In the first way, we compute the time derivative of the radiative
moments $U_{ij}$ and $V_{ij}$ whose SO-tail contributions are given in
Eqs.~\eqref{eq:UVspintail}, and then square these radiative moments to get the
flux \eqref{eq:flux_corrections}. The second way is completely equivalent, but
entirely done by hands. It consists of writing all the separate pieces
composing the energy flux \eqref{eq:flux_corrections}, made of the coupling
between some instantaneous moment (evaluated at current instant $T_R$) times a
hereditary tail integral. The SO terms have to be included in either the
instantaneous moment in front of the integral, or in the tail integral
itself. This gives then several ``direct'' SO contributions coming from tails
at relative 1.5PN order (for the mass quadrupole tail) or 0.5PN order (for the
current quadrupole tail) which are then added together. In addition there is
the crucial contribution due to the reduction to circular orbits of the
standard (non-spin) tail integral at 1.5PN order, for which the relation
between the orbital separation $r$ and the orbital frequency $\omega$ [as
  given by the inverse of Eq.~\eqref{eq:omega}] provides a supplementary SO
term at relative 1.5PN order, which thus contributes \textit{in fine} at the
same 3PN level as the ``direct'' SO tail terms.

Finally, we obtain the
following net result for the SO tail contribution at 3PN order in the total
energy flux:
\begin{align}\label{3PNspintail}
\delta\mathcal{F} &=\frac{32}{5} \frac{c^5}{G}\,x^8\,\nu^2 \Bigl[ - 16 \pi
  \,s_\ell - \frac{31\pi}{6}\,\delta\,\sigma_\ell\Bigr]\,,
\end{align}
where we recall that $s_\ell=\bm{s}\cdot\bm{\ell}$ and
$\sigma_\ell=\bm{\sigma}\cdot\bm{\ell}$, with the spin
variables $\bm{s}$ and $\bm{\sigma}$ being defined by
Eqs.~\eqref{eq:Sbar}--\eqref{eq:ssigma}. Let us remark that in the energy flux
the 3PN SO term is entirely constituted by the SO tails we have obtained in
\eqref{3PNspintail}. So the complete 3PN SO term in the flux is provided by
Eq.~\eqref{3PNspintail}. Contrary to the waveform computed in
Sec.~\ref{sec:comput}, there are no other SO terms coming from linear source
moments at that order.

Because the energy flux and the resulting orbital phasing is so important for
gravitational-wave observations, we shall now give the complete formula for
the total flux, including all non-spin terms and all linear SO terms up to 3PN
order (but neglecting non-linear SS interactions). However we shall not write
the known non-spin 3.5PN terms in the flux (due to non-spin
tails~\cite{B98tail}) because some yet uncalculated SO effects should
conjointly appear at that order. The 3PN energy flux, complete except for SS
interactions, reads then
\begin{widetext}
\begin{align}\label{flux}
  \mathcal{F}=&\frac{32}{5}\frac{c^5}{G}\,x^5\,\nu^2\bigg\{ 1 +x\Bigl(
  -\frac{1247}{336}-\frac{35}{12}\nu \Bigr) +x^{3/2} \Bigl(
  4\pi-4s_\ell -\frac{5}{4}\delta \sigma_\ell \Bigr)
  \nonumber\\&\qquad+x^2 \Bigl(-\frac{44711}{9072}+\frac{9271}{504}\nu
  +\frac{65}{18}\nu^2 \Bigr) \nonumber \\ & \qquad + x^{5/2}
  \Bigl(-\frac{8191}{672}\pi -\frac{9}{2}
  s_\ell-\frac{13}{16}\delta \sigma_\ell +
  \nu\Bigl[-\frac{583}{24}\pi+\frac{272}{9}s_\ell
    +\frac{43}{4}\delta \sigma_\ell\Bigr] \Bigr) \nonumber \\ &
  \qquad + x^{3}\Bigl(\frac{6643739519}{69854400}+
  \frac{16}{3}\pi^2-\frac{1712}{105}\gamma_\text{E} - \frac{856}{105} \ln
  (16\,x) - 16 \pi s_\ell-\frac{31\pi}{6}
  \delta \sigma_\ell \nonumber\\ & \qquad \qquad + \nu
  \Bigl[-\frac{134543}{7776} + \frac{41}{48}\pi^2 \Bigr] -
  \frac{94403}{3024}\nu^2 - \frac{775}{324}\nu^3\Bigr) \bigg\}\,.
\end{align}
\end{widetext}
We are consistently using the constant-magnitude spins $\bm{S}_A$ that are
related to the original variables $\bar{\bm{S}}_A$ of Ref.~\cite{FBB06,BBF06}
by Eq.~\eqref{eq:Sbar}; see also Eqs.~(7.4) of Ref.~\cite{BBF06}. The non-spin
terms are given, e.g., in Ref.~\cite{Bliving}. We find perfect agreement in
the perturbative limit $\nu\rightarrow 0$ with black hole perturbation
calculations reported in Ref.~\cite{TSTS96}. On the other hand the total
conservative energy $E$ of the binary is not affected by the SO terms at the
3PN order (we check this point in Appendix \ref{sec:eomcheck}), hence we have
\begin{widetext}
\begin{align}\label{E}
E =& - \frac{1}{2} m\,\nu\,c^2\,x\bigg\{ 1 +x\Bigl( -\frac{3}{4} -
\frac{\nu}{12} \Bigr) +x^{3/2} \Bigl( \frac{14}{3}s_\ell +
2\delta \sigma_\ell \Bigr) \nonumber\\&\qquad+x^2 \Bigl(
-\frac{27}{8} + \frac{19}{8}\nu -\frac{\nu^2}{24}\Bigr) + x^{5/2} \Bigl( 11
s_\ell + 3\delta \sigma_\ell +
\nu\Bigl[-\frac{61}{9}s_\ell
  -\frac{10}{3}\delta \sigma_\ell\Bigr] \Bigr) \nonumber \\ & \qquad
+ x^{3}\Bigl( -\frac{675}{64} + \left[\frac{34445}{576} - \frac{205}{96}\pi^2
  \right]\nu - \frac{155}{96}\nu^2 - \frac{35}{5184}\nu^3 \Bigr) \bigg\}\,.
\end{align}
\end{widetext}
Following Ref.~\cite{BBF06} we shall next use the standard energy balance
argument to deduce the evolution of the orbital frequency even in the presence
of spins.  To this end we have to check that the constant-magnitude spins are
secularly constant (i.e., constant over a long radiation-reaction time scale)
up to the right level, 1.5PN order in the present case. In Ref.~\cite{BBF06}
we have referred to the work~\cite{W05} for a proof that this is correct up to
relative 1PN order, i.e. considering radiation reaction effects up to 3.5PN
order. In Appendix \ref{sec:secular} below we extend the argument of
Ref.~\cite{W05} to the relative 1.5PN order, which essentially means adding
the tail-induced part of the radiation reaction at 4PN order. This check being
done we can thus neglect $\langle\ud s_\ell/\ud t\rangle$ and $\langle\ud
\sigma_\ell/\ud t\rangle$ in average over a radiation-reaction time scale.

An alternative way to see this is to directly compute the variation of the
projection of the spins along the Newtonian orbital angular momentum,
i.e. $S^A_\ell=\bm{S}_A\cdot\bm{\ell}$, using the precession equations
\eqref{eq:precession_equations} appropriate for constant-magnitude spins. We
readily find that $\ud S^A_\ell/\ud t=\bm{S}_A\cdot[\ud \bm{\ell}/\ud
  t+\bm{\ell}\times\bm{\Omega}_A]$, which shows that $\ud S^A_\ell/\ud t$ is
at least quadratic in the spins for circular orbits. This readily follows from
the facts that $\bm{\ell}$ remains constant in the absence of spins, and that,
as we have seen in Eq.~\eqref{eq:Omega1_circ}, $\bm{\Omega}_A$ for circular
orbits points in the direction of $\bm{\ell}$ modulo spin corrections. Thus we
have $\ud S^A_\ell/\ud t=0$ at the linear SO level (neglecting quadratic SS
couplings). The argument is in principle valid up to any PN order, but is
restricted to circular orbits.

The conclusion is that the constant-magnitude spin terms can be considered as
constant when computing the averaged evolution $\langle\ud E/\ud t\rangle$ of
the energy given by Eq.~\eqref{E}. Equating then $\langle\ud E/\ud t\rangle$
to $-\mathcal{F}$, where $\mathcal{F}$ is given by \eqref{flux}, we obtain the
secular variation of the frequency $\langle\dot\omega\rangle$ --- denoted
$\dot\omega$ for simplicity --- as (neglecting SS contributions)
\begin{widetext}
\begin{align}
\frac{\dot{\omega}}{\omega^2} &= \frac{96}{5}\,\nu\,x^{5/2} \bigg\{ 1 +x
\Bigl(-\frac{743}{336}-\frac{11}{4}\nu \Bigr) + x^{3/2} \Bigl(
4\pi-\frac{47}{3}s_\ell -\frac{25}{4} \delta \sigma_\ell
\Bigr) \nonumber\\&\qquad+x^2 \Bigl( \frac{34103}{18144}+\frac{13661}{2016}\nu
+\frac{59}{18}\nu^2 \Bigr) \nonumber\\&\qquad + x^{5/2} \Bigl(
-\frac{4159}{672} \pi - \frac{5861}{144} s_\ell - \frac{809}{84}
\delta \sigma_\ell + \nu \Bigl[ - \frac{189}{8}\pi +
  \frac{1001}{12} s_\ell + \frac{281}{8} \delta\sigma_\ell
  \Bigr] \Bigr) \nonumber\\ &\qquad + x^{3} \Bigl(\frac{16447322263}{139708800}+
  \frac{16}{3}\pi^2-\frac{1712}{105}\gamma_\text{E} - \frac{856}{105} \ln
  (16\,x) - \frac{188\pi}{3} s_\ell-\frac{151\pi}{6}
  \delta \sigma_\ell \nonumber\\ & \qquad \qquad + \nu
  \Bigl[-\frac{56198689}{217728} + \frac{451}{48}\pi^2 \Bigr] 
+ \frac{541}{896}\nu^2 - \frac{5605}{2592}\nu^3
\Bigr) \bigg\}\,.
\end{align}
\end{widetext}
By integrating this out using standard PN rules for multiplying, dividing and
integrating PN expressions, we obtain the secular evolution of the carrier
phase [defined by $\phi=\int\omega\,\ud t$; see Eq.~\eqref{carrier}] as
\begin{widetext}
\begin{align}\label{carrierphase}
  \phi &= \phi_0 -\frac{1}{32\nu}\,\bigg\{ x^{-5/2} +x^{-3/2} \Bigl(
  \frac{3715}{1008}+ \frac{55}{12}\,\nu \Bigr)+x^{-1}
  \Bigl(-10\pi+\frac{235}{6}s_\ell
  +\frac{125}{8}\delta \sigma_\ell\Bigr) \nonumber \\ &
  \qquad+x^{-1/2} \Bigl(\frac{15293365}{1016064} + \frac{27145}{1008}\,\nu +
  \frac{3085}{144}\,\nu^2 \Bigr) \nonumber \\ & \qquad + \ln x \Bigl(
  \frac{38645}{1344}\pi -\frac{554345}{2016}
  s_\ell-\frac{41745}{448}\delta \sigma_\ell + \nu \Bigl[ -
    \frac{65}{16}\pi-\frac{55}{8}s_\ell
    +\frac{15}{8}\delta\sigma_\ell \Bigr] \Bigr)\nonumber\\ & \qquad
  + x^{1/2} \Bigl(\frac{12348611926451}{18776862720}-
  \frac{160}{3}\pi^2-\frac{1712}{21}\gamma_\text{E} - \frac{856}{21} \ln
  (16\,x) + \frac{940\pi}{3} s_\ell+\frac{745\pi}{6}
  \delta \sigma_\ell \nonumber\\ & \qquad \qquad + \nu
  \Bigl[-\frac{15737765635}{12192768} + \frac{2255}{48}\pi^2 \Bigr] 
+ \frac{76055}{6912}\nu^2 - \frac{127825}{5184}\nu^3\Bigr)\bigg\}\,.
\end{align}
\end{widetext}
We recall that to the carrier phase we have also to add the precessional
correction, arising from the changing orientation of the orbital plane. We
have proved in Eq.~\eqref{Phi} that at the 1PN order the total phase $\Phi$ is
given by $\Phi=\phi-\alpha+\mathcal{O}(c^{-4})$. Thus the precessional
correction is given by $-\alpha$ and is explicitly provided by the
solution \eqref{iotasol}--\eqref{alphasol}. Alternatively, the precessional
correction can be computed numerically~\cite{ACST94}.

\section{Conclusion}
\label{sec:concl}

So far, the search for gravitational waves with LIGO and Virgo detectors has
focused on non-spinning compact binaries~\cite{Abbott:2005kq,Abbott:2007xi,
Abbott:2009qj,Abadie:2010yba,Abadie:2011kd},
although in Ref.~\cite{Abbott:2007ai} single-spin templates were employed, for
the first time, to search for inspiraling spinning compact objects. It is
timely and necessary to develop more accurate templates which include spin
effects. Extrapolating results from the non-spinning case, we expect that, for
maximally spinning objects, reasonably accurate templates would need to be
computed at least through 3.5PN order.

During the last years, motivated by the search for gravitational waves, SO
effects have been computed in the two-body equations of motion through 3.5PN
order~\cite{KWW93,OTO98,TOO01,FBB06,DJSspin,Levi:2010zu,HS11} and energy flux
through 2.5PN order~\cite{K95,OTO98,TOO01,W05,BBF06}. Moreover, SS effects
have been calculated through 3PN order in the conservative
dynamics~\cite{K95,KWW93,MVGer05,Porto06,Porto:2006bt,Hergt:2007ha,SHS07,
  hergt_schafer_08,SHS08,SSH08,Porto:2008tb,PR08,Levi:2010} and multipole
moments~\cite{Porto:2010zg}.

In this paper, building on our previous work~\cite{FBB06,BBF06}, we have
improved the accuracy of the energy flux and gravitational waveform by
computing SO terms induced by tail
effects~\cite{BoR66,CTJN67,HR69,BD86,BD88,BD92,BS93,AF97,B98tail}. Those
effects are due to the back-scattering of linear waves in the curved
space-time geometry around the source. Using the multipolar PN formalism
developed in Refs.~\cite{BD86,BD88,BD92,B95,B98mult}, we have identified and
computed the radiative multipole moments responsible of tail terms involving
SO couplings. More specifically, we have computed those SO tail contributions
to the energy flux at 3PN order and to the gravitational waveforms at 2.5PN
and 3PN order. Those SO tails constitute the complete coefficient at 3PN order
in the energy flux. In particular we find that the energy flux is in complete
agreement with the result of black-hole perturbations in the test-particle
limit~\cite{TSTS96}. Our computation is restricted to quasi-circular
inspiraling orbits, and uses the two-body precessional dynamics at 1.5PN
order.

The computation of SO tail terms in the waveform is summarized in
Sec.~\ref{sec:comput}, and some building blocks and foundation for calculating
tail effects in the PN formalism were reviewed in Sec.~\ref{sec:multipole}.
For the first time, we have computed tail terms when precession effects in the
two-body dynamics are also included. The relevant results for the waveform are
given in Eq.~\eqref{eq:hTTspintail} and in
Appendix~\ref{sec:polarizations}. The SO tail effects in the energy flux and
phasing at 3PN order are given in Sec.~\ref{sec:Eflux}, see in particular
Eqs.~\eqref{3PNspintail}--\eqref{flux} and \eqref{carrierphase}.

Considering the vigorous synergy which is currently taking place between
analytical and numerical relativity for building faithful
templates~\cite{DN09,Buo09,Pan09,Ajith08}, we expect that the results
developed in this paper will help the construction of more accurate analytical
templates describing the entire process of inspiral, merger and ringdown of
black holes in presence of spins.

In the near future we plan to complete the knowledge of SO effects in the
gravitational waveform at 3PN order, by computing the non-tail (i.e.,
instantaneous) SO couplings at 2PN and 3PN orders, and the corresponding $-2$
spin-weighted spherical harmonics (or gravitational modes). This will
constitute a step further with respect to Ref.~\cite{ABFO08} which computed SO
effects in the gravitational modes through 1.5PN order.

\begin{acknowledgments} 
We thank Etienne Racine for fruitful collaboration during the early stages of
this project. A.B. acknowledges support from NSF Grant PHY-0903631 and NASA
grant NNX09AI81G. L.B. acknowledges partial support from Programme
International de Coop\'eration Scientifique (CNRS-PICS).
\end{acknowledgments} 

\appendix

\section{3PN spin terms in the equations of motion}\label{sec:eomcheck}

In this Appendix, we check that there are no SO terms at the 3PN order in the
total conservative invariant energy of the binary given by
Eq.~\eqref{E}. Indeed, we find that the 3PN SO terms in the binary's equations
of motion (say, in harmonic coordinates) can be gauged away. The result is to
be expected because we know that the first SO modification of the radiation
reaction damping force arises at the 4PN order rather than 3PN \cite{W05}.

We compute the near-zone PN metric by solving the Einstein field equations in
harmonic coordinates for the stress-energy tensor \eqref{eq:Tmunu}. We find by
direct PN iteration of the metric, parametrized by means of retarded
potentials $V$, $V_i$, $\cdots$ (see Ref.~\cite{FBB06} for more details), that
the contribution of SO terms at 3PN order in this gauge is given by
\begin{subequations}\label{eq:deltag}
\begin{align} 
\delta g_{00} &= \frac{2 G m_1}{3 c^5} (r_1
\mathop{\dot{a}_1}_\text{S}) +
\frac{4}{3 c^8} \varepsilon_{ijk} S_1^j \ddot{a}_1^i r_1^k   \nonumber \\ &+
\frac{1}{c^8} \text{cst}(t) + 1 \leftrightarrow 2 \, ,\\
\delta g_{0i} & = \frac{4G m_1}{c^4} \mathop{a_1^i}_\text{S} - \frac{10G}{3c^7}
 \varepsilon_{ijk}  S_1^j \dot{a}_1^k  + 1 \leftrightarrow 2 \, ,\\
\delta g_{ij} & = 0 \, ,
\end{align}
\end{subequations}
where we indicate with $\text{cst}(t)$ some irrelevant $\mathcal{O}(c^{-8})$
constant term in space, where we keep the SO parts of the acceleration
un-replaced, and where $1 \leftrightarrow 2$ refers to the same expression but
for particle 2.

The metric \eqref{eq:deltag} yields a 3PN contribution in the equations of
motion of spinning particles which can be calculated from the Papapetrou
\cite{Papa51,Papa51spin} equations of motion (see e.g. Sec.~III in
\cite{FBB06}). The result for the acceleration of particle 1 is
\begin{align}
&\delta a_1^i = \frac{G^2}{c^6 r^4} \varepsilon_{ijk} (m_1 S_2^k - m_2 S_1^k)
  \\ &\quad \quad \times\Big[ \Big(15 (n v)^2 - 3 v^2\Big) n^j + 2 \frac{G
      m}{r} n^j - 6 (n v) v^j \Big]\nonumber\, .
\end{align}
We observe that $\delta a_1^i$ is symmetric under the exchange of particles 1
and 2. A closer inspection reveals that it is in fact given by the second
total time derivative of a certain vector, namely
\begin{equation}
\delta a_1^i = \delta a_2^i = \frac{\ud^2\delta X^i}{\ud t^2} \, ,
\end{equation}
and we find that $\delta \bm{X}=\nu \gamma^3 (\bm{x} \times \bm{\sigma})$ in
the notation of Sec.~\ref{sec:eomspin}.  This is precisely the effect of the
gauge transformation associated with a shift of coordinates $x'^i = x^i +
\delta X^i$. We thus conclude that the 3PN SO terms in the equations of motion
are pure gauge and cannot affect the binary's invariant energy \eqref{E}.

\section{4PN spin secular evolution}\label{sec:secular}

Here we show that the constant-magnitude spins are secularly constant,
i.e. constant over a long radiation-reaction time scale, up to the 4PN order
corresponding to the 1.5PN relative order. In Ref.~\cite{W05} this has already
been proved up to 1PN relative order; here we extend the argument to 1.5PN
order. [In the main text after Eq.~\eqref{E} we present an alternative
  argument valid at any PN order but restricted to circular orbits.]

Following Ref.~\cite{W05} we describe our source by a set of well-separated
extended bodies $A$, supposed to be Newtonian in a first stage. We define
the spin $\mathcal{S}_A^i$ of each of the bodies in the usual Newtonian way
as an integral extending over the volume of body $A$,
\begin{equation}\label{SA}
\mathcal{S}_A^i = \varepsilon_{ijk}\int_A \ud^3\bm{x}\,\rho_*\,(x^j-x_A^j)\,v^k\,,
\end{equation}
where $\rho_*$ denotes the Newtonian (baryonic) mass density, and $x_A^j$ is the
Newtonian center of mass position of the body $A$. At Newtonian level the spin
\eqref{SA} agrees with the definition employed in the present paper. The
``baryonic'' spin~\eqref{SA} is only used for the purpose of this Appendix. The
equation of evolution of the baryonic spin reads as
\begin{equation}\label{dSA}
\frac{\ud \mathcal{S}_A^i}{\ud t} = \varepsilon_{ijk}\int_A \ud^3\bm{x}\,
\rho_*\,(x^j-x_A^j)\,a^k\,.
\end{equation}

The spin precession equation follows from inserting into \eqref{dSA} an
explicit solution for the acceleration in terms of positions and
velocities. The resulting equation is then simplified using some virial
relations appropriate to the case where the compact body is ``stationary'',
see Ref.~\cite{W05}. The secular evolution of the spin is then obtained by
considering the radiation reaction piece of the acceleration in
Eq.~\eqref{dSA}.

At the dominant 2.5PN level the radiation reaction acceleration inside an
isolated body in harmonic coordinates is given by (see, e.g., Ref.~\cite{S83})
\begin{equation}\label{radreac}
a_\text{2.5PN}^i =
\frac{G}{c^5}\biggl[\frac{3}{5}x^jI_{ij}^{(5)}+2\frac{\ud}{\ud
    t}\left(v^jI_{ij}^{(3)}\right) + I_{jk}^{(3)}\,\partial_iU_{jk}\biggr]\,,
\end{equation}
where $I_{ij}$ is the source's STF quadrupole moment (at Newtonian order), and
$U_{ij}$ is the Newtonian potential tensor defined by
\begin{equation}\label{Uij}
U_{ij}(\bm{x},t) = G
\int\ud^3\bm{x}'\rho_*(\bm{x}',t)
\frac{(x^i-x'^i)(x^j-x'^j)}{\vert\bm{x}-\bm{x}'\vert^3}\,.
\end{equation}
(We have $U_{ii}=U$, the usual Newtonian scalar potential.) It can be shown
\cite{W05} that the only contribution at 2.5PN order to the spin precession
equation comes from the velocity-dependent part of Eq.~\eqref{radreac},
i.e.
\begin{equation}\label{radreacdot}
a_\text{2.5PN}^i = \frac{2G}{c^5}\,v^j I_{ij}^{(4)}+\cdots\,.
\end{equation}
The other pieces in the 2.5PN acceleration vanish when the size of the body
tends to zero (compact-body limit) and may be ignored. Using a virial
relation~\cite{W05} we readily obtain
\begin{equation}\label{dSA25PN0}
\left(\frac{\ud \mathcal{S}_A^i}{\ud t}\right)_\text{2.5PN} = 
- \frac{G}{c^5}\,I_{ij}^{(4)}(t)\,\mathcal{S}_A^j \,.
\end{equation}
Because the spin is constant in the lowest approximation the latter result is
a total time-derivative:
\begin{equation}\label{dSA25PN}
\left(\frac{\ud \mathcal{S}_A^i}{\ud t}\right)_\text{2.5PN} = 
\frac{\ud}{\ud t}\left[-
\frac{G}{c^5}\,I_{ij}^{(3)}(t)\,\mathcal{S}_A^j\right]\,,
\end{equation}
which can be moved to the left-hand side and absorbed into a negligible
redefinition of the spin variable at 2.5PN order. When specialized to two
compact bodies the result \eqref{dSA25PN} becomes
\begin{align}\label{eq:delta_S1_dot}
\left(\frac{\ud \mathcal{S}_1^i}{\ud t}\right)_\text{2.5PN} &= 
\frac{\ud}{\ud t} \bigg\{\frac{G^2 m_1 m_2}{c^5 r^2} \biggl[ - 6 (n v) 
(n \mathcal{S}_1) n^i \\&\quad + 4 (n \mathcal{S}_1) v^i + 
4 (v \mathcal{S}_1) n^i - \frac{2}{3} (n v) \mathcal{S}_1^i\biggr] \bigg\}\, .
\nonumber
\end{align}
[Note that the latter expression depends on the specific definition of the
  spin one is using, i.e. in the present case the Newtonian spin defined
  for extended bodies by \eqref{SA}; for the spin variable used in
  \cite{FBB06,BBF06} the expression is expected to be different, but still in
  the form of a total time derivative.]

  The results \eqref{dSA25PN} or \eqref{eq:delta_S1_dot} show that there is no
  secular evolution for the spin at the dominant 2.5PN order (see also
  \cite{Ger00} and references therein for related discussions). Note that
  this conclusion actually applies to any spin variable at dominant order.
  However it has been shown in Ref.~\cite{W05} that in the case of the
  constant-magnitude spin there is also no secular evolution of the spins at
  the next-to-leading 3.5PN order in the radiation reaction. At
  next-to-leading order this result \cite{W05} applies specifically to the
  constant-magnitude spins and uses the radiation reaction acceleration up to
  the 3.5PN order.

We now extend the argument by including the higher-order 4PN correction term
(i.e. 1.5PN radiation-reaction order) which is known to be due to
gravitational wave tails \cite{BD88,B97}. That extension is actually
straightforward since it basically needs only the Newtonian result
\eqref{dSA25PN}. The reason is that including the effect of tails in the
radiation reaction simply amounts to replacing the source Newtonian quadrupole
moment $I_{ij}$ by the tail-corrected expression \cite{B97}
\begin{equation}\label{Iijtail}
I^\text{tail}_{ij}(t) = I_{ij}(t) + \frac{4G M}{c^3}\int_{-\infty}^{t}\ud
t'\,I^{(2)}_{ij}(t')\ln\left(\frac{t-t'}{2\tau_0}\right)\,,
\end{equation}
where $\tau_0$ denotes some arbitrary time scale, for instance the one which
appears in Eqs.~\eqref{eq:tails}. Note the factor $4GM/c^3$ in front of the
tail integral which is twice the factor $2GM/c^3$ in front of the tail
integrals in \eqref{eq:tails}. This factor ensures the consistency between the
work done by the radiation reaction force in the local source and the total
energy flux radiated at infinity from the source \cite{BD92}.

Thus the radiation reaction force including the 4PN tails takes (in harmonic
coordinates) the same form as in Eq.~\eqref{radreac} but with $I_{ij}$
replaced by $I^\text{tail}_{ij}$. This shows that the previous Newtonian
argument still holds for the $\text{2.5PN}+\text{4PN}$ radiation reaction
force and that the effect on the precession equation is still in the form of
some irrelevant total time derivative:
\begin{equation}\label{dSAtail}
\left(\frac{\ud \mathcal{S}_A^i}{\ud t}\right)_{\!\text{2.5PN}+\text{4PN}} =
\frac{\ud}{\ud t}\left[-
  \frac{G}{c^5}\,\mathop{I_{ij}}^{(3)}{}^{\!\!\!\text{tail}}(t)\,\mathcal{S}_A^j\right]\,.
\end{equation}
Hence our conclusion that the constant-magnitude spins are secularly constant
up to 4PN order corresponding to 1.5PN radiation-reaction order.

\section{Gravitational-wave polarizations}\label{sec:polarizations}

We derive in this Appendix the two gravitational-wave polarizations.  They are
computed from the projection formulas \eqref{eq:hp}, using the expression
\eqref{eq:hTTspintail} [together with Eq.~\eqref{eq:hTTspintail_true}] for
$\delta h_{ij}^\text{TT}$. We adopt the convention shown in
Fig.~\ref{figure:SourceFrame} for the polarization vectors. To shorten the
result, we denote the projections of the polarisation basis
$\{\bm{N},\bm{P},\bm{Q}\}$ onto the moving triad $\{\bm{n}, \bm{\lambda},
\bm{\ell}\}$ by e.g. $P_n$, $P_\lambda$, $P_\ell$. With this notation, we have
\begin{widetext}
\begin{subequations}\label{deltahpc}
\begin{align}
\delta h_+ = \frac{G m \nu}{c^2R}& \bigg\{ x^{7/2} \bigg[\pi (2 (P_n Q_\ell
  + P_\ell Q_n) \sigma_\ell - 2 (P_\ell Q_\ell - P_n Q_n + P_{\lambda }
  Q_{\lambda }) \sigma_n + 2 (P_{\lambda } Q_n + P_n Q_{\lambda })
  \sigma_{\lambda }) \nonumber \\ & + 4 \Bigl( - (P_{\lambda } Q_\ell + P_\ell
  Q_{\lambda}) \sigma_\ell - (P_{\lambda } Q_n + P_n Q_{\lambda }) \sigma_n +
  (P_\ell Q_\ell + P_n Q_n - P_{\lambda } Q_{\lambda }) \sigma_{\lambda
  }\Bigr) \Bigl(\ln (2\tau_0\omega )- \frac{7}{6} + \gamma_\text{E} \Bigr)
  \bigg] \nonumber \\ & + x^4 \bigg[ - \frac{16}{3} ( P_n P_{\lambda } + N_n
  P_n Q_\ell - N_{\lambda } P_{\lambda} Q_\ell + N_n P_\ell Q_n + N_\ell P_n
  Q_n - N_{\lambda } P_\ell Q_{\lambda } - N_\ell P_{\lambda } Q_{\lambda } -
  Q_n Q_{\lambda } ) s_\ell \nonumber \\ & - \frac{4}{3} (19 P_\ell P_{\lambda
  } + 12 N_n P_n Q_n - 4 N_{\lambda } P_{\lambda } Q_n - 4 N_{\lambda} P_n
  Q_{\lambda } - 4 N_n P_{\lambda } Q_{\lambda } - 19 Q_\ell Q_{\lambda } )
  s_n \nonumber \\ & - \frac{4}{3} (19 P_\ell P_n + 4 N_{\lambda } P_n Q_n+4
  N_n P_{\lambda } Q_n - 19 Q_\ell Q_n + 4 N_n P_n Q_{\lambda } - 12
  N_{\lambda } P_{\lambda } Q_{\lambda} ) s_{\lambda } \nonumber \\ & + \delta
  \Bigl( - 12 (P_\ell P_{\lambda } - Q_\ell Q_{\lambda}) \sigma_n - 12 (P_\ell
  P_n-Q_\ell Q_n) \sigma_{\lambda }\Bigr) \nonumber \\ & + \pi \Bigl(8
  ({P_n}^2 - {P_{\lambda }}^2 - {Q_n}^2 + {Q_{\lambda }}^2 ) s_\ell +
  \frac{142}{3} (P_\ell P_n-Q_\ell Q_n) s_n - \frac{58}{3} (P_\ell P_{\lambda}
  - Q_\ell Q_{\lambda }) s_{\lambda } \nonumber \nonumber \\ & - \frac{16}{3}
  (N_{\lambda } P_n Q_\ell + N_n P_{\lambda } Q_\ell + N_{\lambda } P_\ell Q_n
  + N_\ell P_{\lambda } Q_n + N_n P_\ell Q_{\lambda } + N_\ell P_n Q_{\lambda
  }) (s_\ell+\delta \sigma_\ell) \nonumber \\ & - \frac{32}{3} (N_{\lambda }
  P_n Q_n + N_n P_{\lambda } Q_n + N_n P_n Q_{\lambda }) (s_n + \delta
  \sigma_n) + \delta \Bigl(\frac{8}{3} ({P_n}^2 - {P_{\lambda }}^2 - {Q_n}^2 +
        {Q_{\lambda }}^2 ) \sigma_\ell \nonumber \\ & + \frac{70}{3} (P_\ell
        P_n - Q_\ell Q_n) \sigma_n-\frac{34}{3} (P_\ell P_{\lambda } - Q_\ell
        Q_{\lambda }) \sigma_{\lambda} \Bigr) - \frac{32}{3} (N_{\lambda }
        P_{\lambda } Q_n + N_{\lambda } P_n Q_{\lambda } + N_n P_{\lambda }
        Q_{\lambda }) (s_{\lambda } + \delta \sigma_{\lambda })\Bigr)
        \nonumber \\ & + \Bigl(-\frac{32}{3} (N_n P_n Q_\ell - N_{\lambda}
        P_{\lambda } Q_\ell + N_n P_\ell Q_n + N_\ell P_n Q_n - N_{\lambda }
        P_\ell Q_{\lambda} - N_\ell P_{\lambda } Q_{\lambda }) (s_\ell +
        \delta \sigma_\ell) \nonumber \\ & - \frac{32}{3} (3 N_n P_n Q_n -
        N_{\lambda } P_{\lambda } Q_n-N_{\lambda } P_n Q_{\lambda } - N_n
        P_{\lambda } Q_{\lambda }) (s_n+\delta \sigma_n) \nonumber \\ & -
        \frac{32}{3} (N_{\lambda } P_n Q_n+N_n P_{\lambda } Q_n + N_n P_n
        Q_{\lambda }-3 N_{\lambda } P_{\lambda } Q_{\lambda }) (s_{\lambda}+
        \delta \sigma_{\lambda })\Bigr) \Bigl(\ln (4\tau_0 \omega ) -
        \frac{5}{3} + \gamma_\text{E} \Bigr) \nonumber \\ & + \bigg(-32 (P_n
        P_{\lambda } - Q_n Q_{\lambda }) s_\ell - \frac{284}{3} (P_\ell
        P_{\lambda } - Q_\ell Q_{\lambda }) s_n - \frac{116}{3} (P_\ell P_n -
        Q_\ell Q_n) s_{\lambda } \nonumber \\ & + \delta \Bigl( - \frac{32}{3}
        (P_n P_{\lambda } - Q_n Q_{\lambda }) \sigma_\ell - \frac{140}{3}
        (P_\ell P_{\lambda } - Q_\ell Q_{\lambda }) \sigma_n \nonumber \\ & -
        \frac{68}{3} (P_\ell P_n - Q_\ell Q_n) \sigma_{\lambda }\Bigr) \bigg)
        \Bigl( \ln (4 \tau_0 \omega ) - \frac{11}{12} + \gamma_\text{E}
        \Bigr)\bigg] \bigg\} \, , \\ 
\delta h_\times = \frac{G m \nu}{c^2
  R} & \bigg\{ x^{7/2} \bigg[\pi \bigg( - 2 (P_\ell P_n - Q_\ell Q_n)
  \sigma_\ell + ({P_\ell}^2 - {P_n}^2 + {P_{\lambda }}^2 - {Q_\ell}^2 +
        {Q_n}^2 - {Q_{\lambda}}^2 ) \sigma_n - 2 (P_n P_{\lambda } - Q_n
        Q_{\lambda }) \sigma_{\lambda}\bigg) \nonumber \\ & + \Bigl(4 (P_\ell
        P_{\lambda } - Q_\ell Q_{\lambda }) \sigma_\ell + 4 (P_n P_{\lambda }
        - Q_n Q_{\lambda }) \sigma_n \nonumber \\ & - 2 ({P_\ell}^2 + {P_n}^2
        - {P_{\lambda }}^2 - {Q_\ell}^2 - {Q_n}^2 + {Q_{\lambda }}^2 )
        \sigma_{\lambda }\Bigr) \Bigl( \ln (2 \tau_0\omega )- \frac{7}{6} +
        \gamma_\text{E} \Bigr)\bigg] \nonumber \\ & + x^4 \bigg[\frac{8}{3} (2
  N_n P_\ell P_n + N_\ell {P_n}^2 - 2 N_{\lambda } P_\ell P_{\lambda } -
  N_\ell {P_{\lambda }}^2 - 2 P_{\lambda } Q_n - 2 N_n Q_\ell Q_n - N_\ell
  {Q_n}^2 - 2 P_n Q_{\lambda } \nonumber \\ & + 2 N_{\lambda } Q_\ell
  Q_{\lambda } + N_\ell {Q_{\lambda}}^2 ) s_\ell + \frac{4}{3} (6 N_n {P_n}^2
  - 4 N_{\lambda } P_n P_{\lambda} - 2 N_n {P_{\lambda }}^2 - 19 P_{\lambda }
  Q_\ell - 6 N_n {Q_n}^2 - 19 P_\ell Q_{\lambda } \nonumber \\ & + 4
  N_{\lambda } Q_n Q_{\lambda } + 2 N_n {Q_{\lambda}}^2 ) s_n + \frac{4}{3} (2
  N_{\lambda } {P_n}^2 + 4 N_n P_n P_{\lambda } - 6 N_{\lambda } {P_{\lambda
  }}^2 - 19 P_n Q_\ell \nonumber \\ & - 19 P_\ell Q_n - 2 N_{\lambda } {Q_n}^2
  - 4 N_n Q_n Q_{\lambda } + 6 N_{\lambda } {Q_{\lambda }}^2 ) s_{\lambda } +
  \delta ( - 12 (P_{\lambda } Q_\ell + P_\ell Q_{\lambda }) \sigma_n - 12 (P_n
  Q_\ell + P_\ell Q_n) \sigma_{\lambda }) \nonumber \\ & + \pi \Bigl(16 (P_n
  Q_n - P_{\lambda } Q_{\lambda}) s_\ell + \frac{142}{3} (P_n Q_\ell + P_\ell
  Q_n) s_n - \frac{58}{3} (P_{\lambda } Q_\ell + P_\ell Q_{\lambda })
  s_{\lambda } \nonumber \\ & + \frac{16}{3} (N_{\lambda } P_\ell P_n + N_n
  P_\ell P_{\lambda } + N_\ell P_n P_{\lambda } - N_{\lambda } Q_\ell Q_n -
  N_n Q_\ell Q_{\lambda } - N_\ell Q_n Q_{\lambda }) (s_\ell + \delta
  \sigma_\ell) \nonumber \\ & + \frac{16}{3} (N_{\lambda } {P_n}^2 + 2 N_n P_n
  P_{\lambda } - N_{\lambda } {Q_n}^2 - 2 N_n Q_n Q_{\lambda }) (s_n + \delta
  \sigma_n) + \delta \Bigl(\frac{16}{3} (P_n Q_n - P_{\lambda } Q_{\lambda })
  \sigma_\ell \nonumber \\ & + \frac{70}{3} (P_n Q_\ell + P_\ell Q_n) \sigma_n
  - \frac{34}{3} (P_{\lambda } Q_\ell + P_\ell Q_{\lambda }) \sigma_{\lambda
  }\Bigr) + \frac{16}{3} (2 N_{\lambda } P_n P_{\lambda } + N_n {P_{\lambda
  }}^2 - 2 N_{\lambda } Q_n Q_{\lambda } \nonumber \\ & - N_n {Q_{\lambda }}^2
  ) (s_{\lambda } + \delta \sigma_{\lambda })\Bigr) + \bigg(\frac{16}{3} (2
  N_n P_\ell P_n + N_\ell {P_n}^2 - 2 N_{\lambda } P_\ell P_{\lambda } -
  N_\ell {P_{\lambda }}^2 - 2 N_n Q_\ell Q_n - N_\ell {Q_n}^2 \nonumber \\ & +
  2 N_{\lambda } Q_\ell Q_{\lambda } + N_\ell {Q_{\lambda}}^2 ) (s_\ell +
  \delta \sigma_\ell) + \frac{16}{3} \Bigl(3 N_n {P_n}^2 - 2 N_{\lambda } P_n
  P_{\lambda } - N_n {P_{\lambda }}^2 - 3 N_n {Q_n}^2 \nonumber \\ & + 2
  N_{\lambda } Q_n Q_{\lambda } + N_n {Q_{\lambda }}^2\Bigr) (s_n + \delta
  \sigma_n) + \frac{16}{3} \Bigl(N_{\lambda } {P_n}^2 + 2 N_n P_n P_{\lambda }
  - 3 N_{\lambda } {P_{\lambda }}^2 - N_{\lambda } {Q_n}^2 \nonumber \\ & - 2
  N_n Q_n Q_{\lambda } + 3N_{\lambda } {Q_{\lambda }}^2\Bigr) (s_{\lambda } +
  \delta \sigma_{\lambda})\bigg) \Bigl( \ln (4 \tau_0 \omega ) - \frac{5}{3} +
  \gamma_\text{E} \Bigr) + \bigg( -32 (P_{\lambda } Q_n + P_n Q_{\lambda })
  s_\ell \nonumber \\ & - \frac{284}{3} (P_{\lambda } Q_\ell + P_\ell
  Q_{\lambda }) s_n - \frac{116}{3} (P_n Q_\ell + P_\ell Q_n) s_{\lambda } +
  \delta \Bigl( - \frac{32}{3} (P_{\lambda } Q_n + P_n Q_{\lambda })
  \sigma_\ell \nonumber \\ & - \frac{140}{3} (P_{\lambda } Q_\ell + P_\ell
  Q_{\lambda }) \sigma_n - \frac{68}{3} (P_n Q_\ell + P_\ell Q_n)
  \sigma_{\lambda }\Bigr)\bigg) \Bigl( \ln (4 \tau_0 \omega )- \frac{11}{12} +
  \gamma_\text{E} \Bigr)\bigg]\bigg\} \, .
\end{align}\end{subequations}
\end{widetext}
The authors can provide on demand a file containing the results in
\textit{Mathematica}${}^\circledR$ input format.


\end{document}